\documentclass[twocolumn]{IEEEtran}
\IEEEoverridecommandlockouts
\usepackage{cite}
\usepackage{amsmath,amssymb,amsfonts}
\usepackage{algorithmic}
\usepackage{graphicx}
\usepackage{subfigure}
\usepackage{textcomp}
\usepackage{color}
\usepackage{tabularx}
\usepackage{array}

\def\BibTeX{{\rm B\kern-.05em{\sc i\kern-.025em b}\kern-.08em
    T\kern-.1667em\lower.7ex\hbox{E}\kern-.125emX}}
    
\begin{document}
\title{Unified Simultaneous Wireless Information and Power Transfer for IoT: Signaling and Architecture with Deep Learning Adaptive Control}
\author{\IEEEauthorblockN{Jong Jin Park, Jong Ho Moon, Hyeon Ho Jang \IEEEmembership{Student Member, IEEE}, and Dong In Kim, \IEEEmembership{Fellow, IEEE}}
\thanks{This research was supported in part by the National Research Foundation of Korea (NRF) Grant funded by the Korean Government (MSIT) under Grant 2021R1A2C2007638.}
\thanks{The authors are with the Department of Electrical and Computer Engineering, Sungkyunkwan University (SKKU), Suwon 16419, South Korea (Email: \{pjj0805, moonjh525, jhh3727\}@skku.edu; dikim@skku.ac.kr).}
}

\maketitle
\begin{abstract}
In this paper, we propose a unified SWIPT signal and its architecture design in order to take advantage of both single tone and multi-tone signaling by adjusting only the power allocation ratio of a unified signal. For this, we design a novel unified and integrated receiver architecture for the proposed unified SWIPT signaling, which consumes low power with an envelope detection. To relieve the computational complexity of the receiver, we propose an adaptive control algorithm by which the transmitter adjusts the communication mode through temporal convolutional network (TCN) based \textit{asymmetric processing}. To this end, the transmitter optimizes the modulation index and power allocation ratio in short-term scale while updating the mode switching threshold in long-term scale. We demonstrate that the proposed unified SWIPT system improves the achievable rate under the self-powering condition of low-power IoT devices. Consequently it is foreseen to effectively deploy low-power IoT networks that concurrently supply both information and energy \textit{wirelessly} to the devices by using the proposed unified SWIPT and adaptive control algorithm in place at the transmitter side.
\end{abstract}

\section{Introduction}
The Internet of Things (IoT) has emerged with the ever growing interests in the large-scale deployment of low-energy devices that collect information while connecting billions of devices through networks. But powering a huge number of IoT devices is expected to be a significant challenge in the upcoming IoT era. Furthermore, the limited battery lifetime is an important issue for permanently operating low-energy IoT networks. Therefore, instead of battery replacement, self-powered devices charged from ambient renewable resources could be deployed to self-sustain the IoT networks. 

Recently, thanks to the reduction in power consumption of device circuits, and with the need to energize low-power devices, radio frequency (RF) wireless power transfer (WPT) has attracted attention as a promising power supply technology for self-powering IoT devices. By utilizing the property that RF signal can convey both information and power via the same electromagnetic (EM) wave, the concept of WPT can be extended to simultaneous wireless information and power transfer (SWIPT) \cite{TDP}. It is suitable for a system that supports low-power wireless devices or sensors, such as IoT networks. There have been many proposals \cite{XZ,BCfun} that aim at enhancing the rate-energy tradeoff through optimum receiver design for SWIPT. Two of the most widely used SWIPT schemes, time switching (TS) and power splitting (PS) have been studied in \cite{XZ}. PS is based on the division of the signal power into two streams for energy harvesting (EH) and information decoding (ID). In TS, the received signal is used for EH or ID in specific time periods. In \cite{SAT}, the authors studied the performance of the integrated receiver (IntRx) utilizing a rectifier for EH and ID while reducing the power consumption by integrating EH and ID into one common circuit.

Meanwhile, it was shown that the RF to direct current (RF-to-DC) power conversion efficiency (PCE) in EH circuits of SWIPT receivers depends not only on the input power but also the shape of the input signal, i.e., peak-to-average power ratio (PAPR), due to the nonlinearity of the rectifier, such as diode small signal \cite{BCwpt, RM} and saturation effect \cite{EB}. Following this observation, new modulation schemes which utilize multi-tone waveforms for boosting the PCE were proposed in \cite{DIK,IK,SC,TI}. The authors in \cite{DIK} proposed the PAPR modulation for SWIPT using distinct levels of signal PAPR to convey information. Similarly, the authors in \cite{IK} designed tone-index multisine modulation where information is embedded into the tone-index of multi-tone. These multi-tone SWIPT schemes utilize an envelope detection that requires less power consumption and low complexity for ID since the receiver can be implemented without power-hungry devices (e.g., mixer, {\em I/Q} demodulator). Also, the authors in \cite{SC,TI} proposed multi-tone FSK based SWIPT by using the relationship between the input frequency spacing and rectifier output intermodulation frequencies. But most of the multi-tone SWIPT schemes suffer lower data rate compared to conventional information modulation, hence the rate-energy tradeoff should be optimized jointly with the operational range of IoT devices.

To assess the power transfer efficiency of SWIPT accurately, the recent works \cite{HHJ, PM} highlighted the nonlinear impact of the high power amplifier (HPA) to SWIPT, in which the HPA can significantly degrade the performance of the conventional SWIPT. Furthermore, multi-tone signals with high PAPR are more sensitive to the nonlinearity of HPA which can distort the amplitude of signal. Such distortion on signal waveform significantly impacts on both EH and ID performance of multi-tone SWIPT, as studied in \cite{JJPlet}. Also, the high PAPR signal can cause in-band interference to nearby unintended information receivers, such as low noise amplifier (LNA) saturation \cite{BAM}. To avoid such interference, the authors in \cite{KWC} proposed a frequency-splitting (FS) architecture for SWIPT which sends power via the unmodulated high-power continuous wave (CW) and transmits information by using a small modulated signal with orthogonal frequency-division multiplexing (OFDM). 

Consequently the effect of the transmitter HPA on signal waveforms should be jointly considered in conjunction with the receiver rate-energy tradeoff optimization, which leads to the end-to-end efficiency of SWIPT systems. Therefore, a new modulation scheme and an integrated receiver architecture optimized for both the transmitter HPA and the receiver rectifier nonlinearity are required to maximize the end-to-end efficiency of SWIPT systems, thereby realizing the self-powered IoT networks. This has motivated our work. 

In cellular networks, we utilize a link adaptation (adaptive modulation and coding) scheme depending on channel quality and required quality of service (QoS). In the same spirit, in low-energy IoT networks, we may control the transmit power and communication mode considering the received power of self-powered devices via SWIPT. But the end-to-end efficiency of SWIPT is highly nonlinear due to nonlinear devices (i.e., transmitter HPA and receiver rectifier). Hence the optimization over such nonlinearity is not as simple as conventional look-up table (LUT) based link adaptation \cite{3GPP}. 

To tackle this difficulty, machine learning (ML) has recently been introduced into wireless communications, so as to handle various nonlinear and complex problems of the communication systems by extracting the inherent features from data \cite{CZ}. The ML enables the network infrastructure to learn from the environment and take adaptive network optimization \cite{MC}. For example, the authors in \cite{FT} have proposed deep learning based traffic load prediction to predict the network congestion and adaptively assign channel to IoT devices.  In \cite{JKlet,QB,JK}, various deep learning methods were proposed to estimate the nonlinear channel characteristics (e.g., time-varying) for adaptive network optimization. Also, \cite{CC} has proposed the reinforcement learning based adaptive control for SWIPT. But these ML methods require parallel computing devices (e.g., GPU) to train models and make inferences, which is a heavy burden on low-energy IoT devices. For this reason, the transmitter-oriented asymmetric processing for adaptive mode switching (MS) was proposed via the recurrent neural network (RNN) with long-short term memory (LSTM) to extract the features from temporal correlation of the channel \cite{JJP}.


Meanwhile, recent works reveal that convolutional neural network (CNN) architectures can outperform the conventional RNNs \cite{SB}. Especially, the temporal convolutional network (TCN) architecture outperforms the recurrent architectures across various sequence modeling tasks \cite{CL,YAF}. Moreover, the overall framework of TCN is simpler than LSTM, and it has natural superiorities for computational advantage of CNNs which include parallelism operation. Motivated by this, the authors in \cite{YC} have proposed a semi-supervised TCN model for anomaly detection in IoT communication. But, to the best of our knowledge, adopting these superiorities of TCN for adaptive control of IoT networks has not been studied in the literature. This also motivated our work. 

In this paper, we propose a novel unified SWIPT signaling and its architecture which utilize both single tone and multi-tone shaped signals in one unified and integrated receiver. Compared to the conventional SWIPT, the proposed SWIPT architecture can handle both the HPA and rectifier nonlinearity and adaptively control the communication mode through the TCN based asymmetric processing. Furthermore, the proposed signal and architecture can be low-power and low-complexity by virtue of the EH - ID dual operation of an envelope detector. The main contributions of this work are highlighted as: 
\begin{itemize}
\item We propose a new unified signal design which can be used for both single tone and multi-tone SWIPT signal transmission. The MS between single tone and multi-tone can be triggered by adjusting only the power allocation ratio of the unified signal. Since the power allocation is done at the transmitter side, the receiver does not need any additional optimization for MS between ID and EH, unlike conventional TS or PS SWIPT. 

\item For the proposed unified SWIPT signal, we also design a novel unified receiver architecture which is the hybrid of FS and PS, but it requires low-power consumption thanks to the envelope detection. Since the unified receiver can operate regardless of the signal waveform shape (i.e., no MS at the receiver), it is suitable for implementing low-energy IoT devices. Using the proposed unified SWIPT signal and architecture design, we analyze the symbol-error rate (SER) and outage performance.

\item In order to lower the computational burden of the receiver, the MS control based on asymmetric processing through TCN is introduced, by which the transmitter optimizes the modulation index and power allocation ratio based on the received power feedback from the receiver. The proposed algorithm determines the communication mode and modulation index over each short-term channel block to meet the energy-causality constraint for self-powering IoT devices. The long-term optimization of the control algorithm takes the short-term optimization results (i.e., MS attributes) as the input data of TCN. The proposed mixed-time scale iterative algorithm effectively controls the unified SWIPT system by estimating the MS threshold from the MS attributes and temporal correlation of time-varying channel.

\item Taking into account the nonlinearities of the transmitter HPA and the receiver rectifier, the proposed adaptive control algorithm adjusts the communication mode and modulation index to optimize the rate-energy tradeoff, considering the end-to-end efficiency. We confirm that the proposed unified SWIPT system with adaptive control improves the performance over the existing SWIPT, under the self-powering condition of IoT devices. 
Therefore, the proposed unified SWIPT with adaptive control algorithm will facilitate implementing low-power IoT networks that simultaneously supply information and energy {\it wirelessly} to the devices.
\end{itemize}

The rest of the paper is organized as follows. In Section II, we describe the unified SWIPT system architecture along with the unified signaling model, and provide examples of the unified receiver operation for ID and EH from the signal. In Section III, we propose the TCN based adaptive control algorithm for maximizing the achievable rate of the unified SWIPT system under the energy-causality constraint to self-sustain the low-energy IoT devices. Section IV presents the simulation results to evaluate the performance of the proposed unified SWIPT with adaptive control algorithm. Then, Section V concludes the work.

\section{System Model}
\begin{figure}
\centering
\includegraphics[width=3.4in]{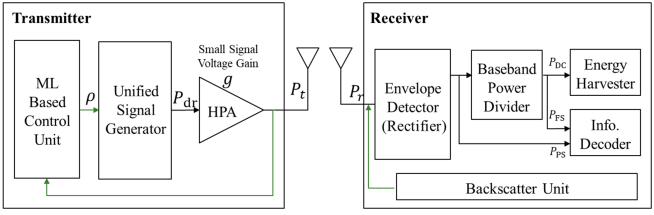}
\caption{Proposed architecture of unified SWIPT transceiver.}
\label{fig:TxRx}
\end{figure}

Fig.~\ref{fig:TxRx} illustrates the proposed architecture of unified SWIPT transceiver. We consider a point-to-point SWIPT system, each equipped with a single antenna, while our design can be applied to the multi-antenna case. Let $h_v$ denote the complex channel gain in the $v$th fading block with bandwidth $B$ and block time $T<T_c$ for the channel coherence time $T_c$. 

For a unified SWIPT signal transmission, the transmitter generates a {\it unified} signal for a given power allocation ratio $\rho$. The power allocation ratio determines overall shape of the unified signal waveform, which will be elaborated in Section \ref{sec:UnifiedSignalDesign}. We here employ PAPR modulation \cite{DIK} for enabling the low-power ID and boosting the PCE at low received power, in which the modulation index of the unified SWIPT is defined as the number of multi-tones $Q$. Then, the PAPR modulated unified signal is amplified by the HPA at the transmitter. 

At the receiver, the received signal is downconverted into its baseband signal by an envelope detector, which is then used for both EH and ID. Note that the receiver for the proposed SWIPT signaling has one {\it unified} receiver for not only EH/ID integration (i.e., the integrated receiver \cite{XZ}) but also single tone/multi-tone SWIPT dual mode operation. The received signal is detected by the envelope detector, and then its baseband signal is split with ratio $\rho_r$ by the power splitter. After the power splitter, the $\rho_r$-portion of the baseband signal, i.e., $P\textsubscript{PS}$ is used for ID, namely {\it power-splitting} (PS) PAPR ID. 
Meanwhile, the $(1-\rho_r)$-portion of the baseband signal is further divided into direct current (DC) power $P\textsubscript{DC}$ and alternating current (AC) power $P\textsubscript{FS}$ by the baseband power divider, in a {\it frequency-splitting} (FS) manner \cite{KWC}. Then $P\textsubscript{FS}$ is selectively used for FS PAPR ID at the information decoder. Consequently $P\textsubscript{DC}$ is used for EH at the energy harvester to assure the self-powering of the receiver. 

With the above operation at the transceiver, the transmitter selects a waveform according to the proposed adaptive control algorithm. 
The ML based control unit at the transmitter updates the MS threshold $P\textsubscript{th}$ and selects the proper power allocation ratio $\rho$ in each fading block, considering the received power $P_{r,v}$ that is measured and reported by a target IoT device in the $v$th fading block. Here, the receiver feedbacks the measurement using the monostatic backscatter \cite{NVH} prior to each frame transmission, so as to train the deep learning network and estimate the channel gains (i.e., CSIT is assumed).\footnote{As the single tone (continuous wave) pilot signal from the transmitter is mainly used for EH, its signal power is much higher than the sensitivity of the monostatic backscatter. Hence, we assume that the signal-to-noise ratio (SNR) of the backscatter always satisfies the required QoS for uplink feedback transmission, assuring {\it error-free} feedback.} In each fading block, all computational overhead for the adaptive control algorithm is shifted to the transmitter, so that the receiver can operate with low power and low complexity, thereby realizing \textit{transmitter-oriented asymmetric processing}. 
We will elaborate more on the control of the system with feedback in Section \ref{sec:TCNalgorithm}. 

\subsection{Unified Signal Design for SWIPT with HPA}
\label{sec:UnifiedSignalDesign}
\begin{figure}
\centering
\includegraphics[width=3.2in]{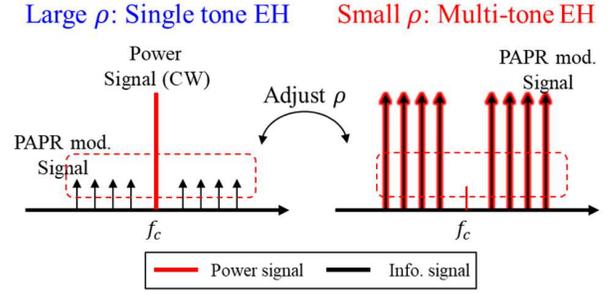}
\caption{Power allocation of unified SWIPT signal and corresponding single tone/multi-tone mode signal spectrum in frequency domain.}
\label{fig:Signal}
\end{figure}

As illustrated in Fig.~\ref{fig:Signal}, we propose a new unified SWIPT signaling which is composed of both single tone CW signal at the carrier frequency $f_c$ and PAPR modulated multi-tone signal occupying some bandwidth around the single tone. By adaptively adjusting the power allocation ratio $\rho$, the unified SWIPT not only mitigates signal distortion from the HPA but also performs single tone/multi-tone MS to achieve better PCE for self-powering. On the one hand, a large $\rho$ means that the transmitter allocates most of the signal power to the single tone at the carrier frequency, which exploits the higher PCE with single tone at the receiver. On the other hand, as $\rho$ decreases, the power allocation to the multi-tone is increased and the overall signal will be shaped in multi-tone, which brings in the favorable PCE with multi-tone at the receiver. Therefore, it allows to adaptively switch the single tone/multi-tone mode by changing only the power allocation ratio using the unified SWIPT signaling above. Especially, we can perform the power allocation required for EH and ID at the transmitter, which is far less complex than conventional TS or PS SWIPT schemes that adjust the power allocation between EH and ID through mathematical optimization at the receiver. 
 
The PAPR based information transmission facilitates low-power decoding via simple PAPR measurements which does not require any power-hungry device (e.g., mixer) \cite{DIK}. 
With the power allocation ratio $\rho$ and the HPA drive power $P\textsubscript{dr}$, the complex-valued baseband signal (i.e., complex envelope) of the PAPR modulated unified SWIPT, which uses a subset of $N$ tones from $Q$ available tones (i.e., $N\in\{1,...,Q\})$, can be expressed as 
\begin{equation}
\begin{aligned}
s(t) &= s_{s}(t) + s_{m}(t) = \sqrt{2\rho P\textsubscript{dr}}\, s_c \exp(j \theta_c)\\
&+\sqrt{2(1-\rho)P\textsubscript{dr}}\,\sum_{n=1}^{N} s_n \exp \big[\!\: j(2\pi  f_{n} t+\theta_{n}) \!\:\big], \,\, 0\leq t \leq T
\end{aligned}
\end{equation}
where $f_n = f_1 +(n-1)\Delta f$ with minimum tone spacing $\Delta f$. To achieve the maximum baseband PAPR of $s_m(t)$, the amplitude and phase of each tone is  $s_c=1$, $s_{n}=\sqrt{1/N}$, and $\theta_c = \theta_n = 0$. Then, the baseband PAPR modulated signal has the form 
\begin{equation}
\begin{aligned}
s(t) = \sqrt{2\rho P\textsubscript{dr}}
+\sqrt{\frac{2(1-\rho)P\textsubscript{dr}}{N}}\!\:\sum_{n=1}^{N} \exp \left(j2\pi  f_{n} t\right).
\end{aligned}
\end{equation}
In the above, we maintain the same peak value as in passband signal, but the average power of complex envelope is $2 P\textsubscript{dr}$. 
Hence, the PAPR of the complex envelope is given by $N$. 

Under the CSIT assumption, precoding (matched filtering) is performed to yield the maximum PAPR at the receiver. With the estimated complex channel gain $g_n$, we define $\mathbf{g}_n=[ g_1,\ldots ,g_N]$ for multi-tone and $g_{c}$ for single tone. Then, the amplitude and phase of each tone is set to $s_c \exp{(j \theta_c)}=g^*_{c}/|g_{c}|$ and 
\begin{equation}
 s_{n}\exp\left({j\theta_{n}}\right)=\sqrt{\frac{1}{N}} \frac{g^*_{n}}{\|\mathbf{g}_{n}\|} 
\end{equation}
where $(\cdot)^*$ denotes complex conjugate. Hence, the passband signal after precoding can be expressed as 
\begin{equation}
\begin{aligned}
x(t)  &=\sqrt{2\rho P\textsubscript{dr}}\!\:\cos(2\pi f_c t + \theta^*_c) \\
	&\quad \qquad+\sqrt{\frac{2(1-\rho)P\textsubscript{dr}}{N}}\!\:\sum_{n=1}^{N}  \cos \big[\!\: 2\pi (f_c +f_{n}) t + \theta^*_n \!\:\big] \\
	&=\text{Re}\big\{ s(t) e^{ j2\pi f_c t} \big\}
\end{aligned}
\end{equation}
where $\theta^*_c = \angle\!\: g^*_{c}/|g_{c}|$ and $\theta^*_n=\angle\!\: g^*_{n}/\|\mathbf{g}_{n}\|$. 
Moreover, if channel is frequency flat (FF), namely $g_{c}=g_n=h$ for $\forall{n}$ (we omit the channel index $v$ for simplicity), then the passband signal can be simplified to
\begin{equation}
\label{eqn:passband}
\begin{aligned}
x(t)&=\text{Re}\Bigg\{ \frac{h^*}{|h|} \Bigg[\sqrt{2\rho P\textsubscript{dr}} \\
&\qquad \qquad \qquad+\sqrt{\frac{2(1-\rho)P\textsubscript{dr}}{N}}\!\:\sum_{n=1}^{N} e^ {j2\pi  f_{n} t}\Bigg]e^{j2\pi f_c t}\Bigg\}\\
&=\text{Re}\left\{ \frac{h^*}{|h|}\!\: s(t)\!\: e^{j2\pi f_c t}\right\} \\
&= \text{Re}\left\{|s(t)|\!\: e^{j\big[ \theta_s(t) + \theta\textsubscript{MF}\big]}  e^{j2\pi f_c t}\right\} 
\end{aligned}
\end{equation}
where $\mathbb{E} [|x(t)|^2]=P\textsubscript{dr}$, $\theta_s(t)$ and $\theta\textsubscript{MF}=\angle\!\: h^*/|h|=-\angle\!\: h$ denote the phase of $s(t)$ and the pre-matched filtering in FF fading channel, respectively.

\begin{figure}
\centering
\includegraphics[width=3.6in]{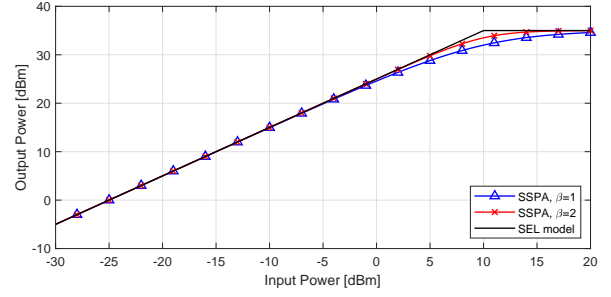}
\caption{Output versus input power based on SSPA ($\beta=1, 2$) and SEL models where data was obtained from Mini Circuits ZHL-5W-422+ HPA.}
\label{fig:HPA}
\end{figure}

The passband signal $x(t)$ is amplified by the HPA after the signal generator, which causes nonlinear amplitude and phase distortions. We assume that the HPA is memoryless, and the complex envelope of the HPA input signal can be expressed by using the polar form $\tilde x\textsubscript{in}(t)= A(t)e^{j\phi(t)}$. Thus, for the unified SWIPT signal with PAPR modulation, the HPA input signal is given by $A(t)=|s(t)|$ with $\phi(t)=\theta_s(t) + \theta\textsubscript{MF}$. The complex envelope of  the HPA output signal is generally written as 
\begin{equation}
\tilde x\textsubscript{out}(t)= G[A(t)]e^{j\{\phi(t)+\Phi[A(t)]\}}
\end{equation}
where $G(\cdot)$ and $\Phi(\cdot)$ represent the amplitude-to-amplitude modulation (AM/AM) and amplitude-to-phase modulation (AM/PM) conversion of the amplifier, respectively. 
To account for the nonlinearity of HPA, the solid-state power amplifier (SSPA) model \cite{JQ} is adopted, which is expressed as
\begin{equation}
G(\mathcal A)=\frac{g\mathcal A}{\left[1+\left(\frac{\mathcal A}{A\textsubscript{sat}}\right)^{2\beta} \right]^\frac{1}{2\beta}} \quad \text{and} \quad \Phi(\mathcal A)=0. 
\end{equation}
Here, $g$ is the amplifier small signal voltage gain, $A\textsubscript{sat}$ is the input saturation level, and $\beta$ is a control parameter to adjust the AM/AM sharpness of the transition from linear region to saturation region. Fig.~\ref{fig:HPA} shows the input-output power characteristics for the considered SSPA model. Note that as $\beta$ increases, the nonlinear transition region before saturation is linearized, which is the same as the soft envelope limiter (SEL) HPA model \cite{JQ}. 


With the above SSPA model, the transmitted passband signal after the HPA is of the form 
\begin{equation}
\begin{aligned}
\label{eqn:HPAout}
 \hat x (t)&=\text{Re}\big\{\tilde{x}\textsubscript{out}(t) e^{j2\pi f_c t} \big\} \\
 &=\text{Re}\Big\{G\left(|s(t)|\right)e^{j\big[ 2\pi f_c t+\theta_s(t) +\theta\textsubscript{MF} \big]}\Big\}
 \end{aligned}
\end{equation}
with the complex envelope $\tilde x\textsubscript{out}(t) =G\left(|s(t)|\right)e^{j\big[\theta_s(t) +\theta\textsubscript{MF}\big]}$.
The average power of the HPA output signal is evaluated as
\begin{equation}
\mathbb{E} \big[|\hat x(t)|^2\big]=\frac{1}{T}\int_{T}|\hat x(t)|^2 dt=\frac{1}{2T}\int_{T}\frac{g^2A\textsubscript{sat}^2 |s(t)|^2}{\big[A\textsubscript{sat}^{2\beta}+|s(t)|^{2\beta}\big]^\frac{1}{\beta}} dt.
\end{equation}
It is challenging to obtain the closed-form expression of the average power with HPA above. In general, the average power of multi-tone shaped signal is a function of the number of tones $N$ due to the HPA distortion \cite{JJPlet}. On the other hand, the average power of single tone shaped signal depends on $\rho$. With $\rho \approx 1$, the average power becomes nearly constant with some compression due to the HPA nonlinearity. As $\rho$ decreases, however, the PAPR of the unified SWIPT signal increases due to the multi-tone portion, which incurs the same distortion as in the multi-tone case. Consequently the average power of the unified SWIPT signal can be expressed as a function of both $\rho$ and $N$, i.e., $\mathcal P (\rho, N)$. Note that the average power of the transmitted signal remains constant to be $\mathcal P(\rho, N)=g^2P\textsubscript{dr}$, regardless of $\rho$ and $N$, in case of ideal amplifier. 

\subsection{Unified Receiver with PAPR Modulation: How It Works}
\label{sec:UnifiedReceiver}
\begin{figure}
\centering
\subfigure[\hspace{-0.4in}]{
\includegraphics[width=3.2in]{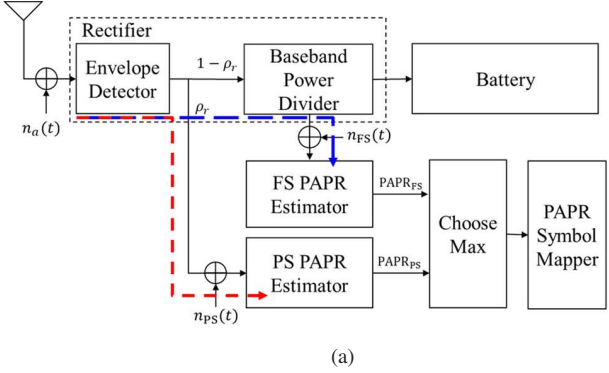}
}
\subfigure[]{
\includegraphics[width=3.6in]{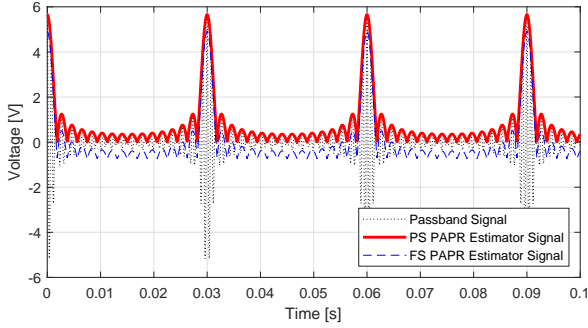}
}
\subfigure[]{
\includegraphics[width=3.6in]{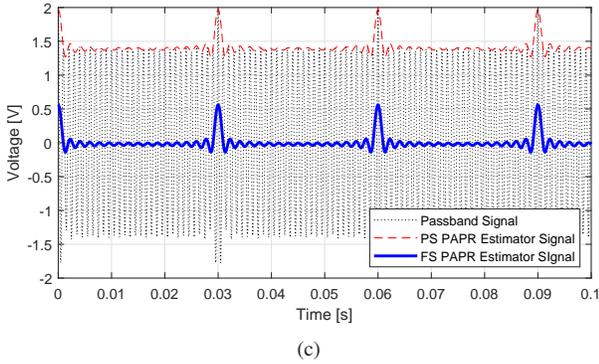}
}
\caption{Proposed unified SWIPT receiver and signal detection of single tone/multi-tone waveforms. (a) Modeling of the unified SWIPT receiver. (b) Multi-tone signal waveform in time domain. (c) Single tone signal waveform in time domain. In (b) and (c), we normalized the signal power (i.e., the power-splitting ratio $\rho_r$ is ignored) for each PAPR estimator to compare the received passband and baseband signals easily.}
\label{fig:Rx}
\end{figure}

In this section, we analyze the {\it dual mode} signal detection of the proposed unified SWIPT receiver as shown in Fig.~\ref{fig:Rx}, which can detect both single tone and multi-tone signals. For this, we employ two PAPR estimators for FS PAPR ID and PS PAPR ID. To simplify the receiver architecture, we may use one PAPR estimator by having a switch between the red and blue signal paths in Fig.~\ref{fig:Rx}(a). But this switching-based receiver architecture requires additional MS information or optimization at the receiver, so as to identify the type of the signal waveform (i.e., single tone or multi-tone) for switching to the proper signal path. Note that the main purpose of the unified receiver is to detect both single tone and multi-tone signals without additional optimization at the receiver (e.g., TS, or PS of \cite{XZ}). Also, the circuit power consumption of the PAPR estimator is little compared to the conventional {\em I/Q} demodulator. Thus, using two PAPR estimators does not cause any burden on the power consumption at the receiver.

At the receiver, the received signal after antenna is expressed as
\begin{equation}
\begin{aligned}
 y (t)&=\text{Re}\big\{ h\!\:\tilde x\textsubscript{out}(t) e^{j2\pi f_c t} \big\}+n_a(t) \\
 &=|h|\!\: G\left(|s(t)|\right) \cos\big[ 2\pi f_c t + \theta_s (t) \big] + n_a(t) 
\end{aligned}
\end{equation}
where $n_{a}(t)=\sqrt{2}\,\text{Re}\big\{\tilde{n}_a(t) e^{j2\pi f_c t}\big\}$ is the antenna noise and $\tilde{n}_a(t)$ is modeled to be a circularly symmetric complex Gaussian (CSCG) noise with $\tilde{n}_a(t)\sim\mathcal{C}\mathcal{N}(0,\sigma_a^{2})$.

\begin{figure}
\centering
\includegraphics[width=3.6in]{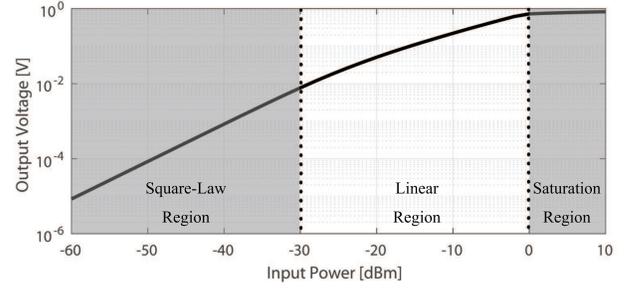}
\caption{Output voltage versus input power of a simple diode detector. We used the Skyworks SMS7630 Schottky diode, $R=1$k$\Omega$ and $C=22$pF with matched network at -10dBm input power.}
\label{fig:Diode}
\end{figure}

For analysis of the unified receiver, we first consider the operating region of a diode detector. The response of the diode detector is generally indicated as a curve of detected output voltage versus input power as shown in Fig.~\ref{fig:Diode}, where the detector circuit is composed of diode, parallel capacitor and resistor. At small input signal, the output voltage is closely proportional to the input power (i.e., the square of the input voltage). This is called the square-law region. In this region, a 10 dB increase of input power results in a 10 times increase of output voltage. On the other hand, the detector output voltage is directly proportional to the input voltage at large input signal. This corresponds to the linear region where a 10 dB increase of input power results in almost 5 times increase of output voltage. At very large input signal, the output voltage is saturated due to the breakdown voltage of the diode. This region is called the saturation region. Since the envelope detector of the proposed receiver is used for both EH and ID, the receiver requires sufficient input power enough to charge the battery for self-powering at the unified SWIPT receiver. This implies that the ID component can be activated only when the harvested power from the input signal satisfies the minimum requirement of power consumption at the PAPR estimator. Therefore, we assume that the operating range of the envelope detector lies in the linear region. 

With the linear operating assumption of the diode envelope detector above, it can track the real envelope of the signal which yields the envelope detector output as
\begin{equation}
y\textsubscript{env}(t) = \big| |h|G\left(|s(t)|\right)+n_a(t) \big|. 
\label{envout}
\end{equation}
We assume that the noise of the diode envelope detector (rectifier) and the following PAPR estimator is much larger than antenna noise (i.e., thermal noise), thus we ignore the antenna noise. 
After the envelope detector, the power splitter divides the baseband signal with ratio $\rho_r$.\footnote{Since the conventional dynamic power-splitting (DPS) scheme requires additional optimization at the receiver, we use the static power splitting (SPS) to reduce complexity of the unified receiver. Also, our unified receiver can be regarded as an integrated receiver due to the baseband PS, which requires small portion (e.g., $\rho_r=10^{-3}$) of the received signal power for ID \cite{XZ}.} Then the signal at the PS PAPR estimator is given by 
\begin{equation}
y\textsubscript{PS}(t) = \sqrt{\rho_r}\, y\textsubscript{env}(t) +n\textsubscript{PS}(t) =  \sqrt{\rho_r}\!\: |h|G\left(|s(t)|\right) + n\textsubscript{PS}(t)
\end{equation}
for the average power $\mathcal P\textsubscript{PS}(\rho,N) = 2\rho_r|h|^2\mathcal P(\rho,N) + \sigma\textsubscript{PS}^{2}$ and the PS PAPR estimator noise $n\textsubscript{PS}(t) \sim\mathcal{N}(0,\sigma\textsubscript{PS}^{2})$. Note that the average power of the signal is doubled due to the imaginary part of the complex-valued baseband signal. Because the diode envelope detector tracks the envelope of the passband signal, the extracted envelope is equal to the magnitude of the complex envelope.

Assuming a linear HPA, and ignoring the noise, the PAPR of PS path is evaluated as 
\begin{equation}
\begin{aligned}
\text{PAPR}\textsubscript{PS}(N)&=2\!\:\frac{\max_{t}|y\textsubscript{PS}(t)|^2}{\mathbb{E}\big[ |y\textsubscript{PS}(t)|^2 \big]}\\
&= 2\!\:\frac{\big[ \sqrt{2 \rho P\textsubscript{dr}}+\sqrt{2(1-\rho)N P\textsubscript{dr}} \,\big]^2}
{2\rho P\textsubscript{dr} + 2(1-\rho)P\textsubscript{dr}}\leq 2N.
\end{aligned}
\end{equation}
Note that the maximum PAPR value of the complex envelope at the PS PAPR estimator is $N$, unlike the passband PAPR of $2N$. For scaling and fair comparison with the FS PAPR later, we simply multiplied the PAPR value by 2. Therefore, the PAPR value at the PS PAPR estimator is $\text{PAPR}\textsubscript{PS}\leq2N$ and equality holds when $\rho=0$, which is multi-tone mode. This result means that the PS PAPR estimator is suitable for multi-tone shaped symbol demodulation. As shown in Figs.~\ref{fig:Rx}(b) and \ref{fig:Rx}(c), we can observe that the detected signal at the PS PAPR estimator (red line of each figure) tracks the envelope of the received passband signal. In Fig.~\ref{fig:Rx}(c), the detected envelope of single tone shaped signal with large $\rho$ at the PS PAPR estimator is biased with some DC component which is downconverted from CW at $f_c$. Thus, the PAPR value of single tone mode signal turns out to be small as the difference between the peak and average values is small. 

After the power splitter, $(1-\rho_r)$-portion of the signal is fed to the baseband frequency power divider (i.e., LC circuit which separates DC and AC components of the baseband signal), where the series capacitance $C_f$ and shunt inductance $L_f$ of LC circuit blocks the DC component of the signal, as shown in Fig.~\ref{fig:circuit}(a). Using the Fourier transform pair $y\textsubscript{env}(f)$ of the envelope signal in (\ref{envout}), the frequency-domain expression of the input signal at the FS PAPR estimator is given by 
\begin{equation}
\label{eqn:fDomainSignal}
y\textsubscript{FS}(f) = \sqrt{1-\rho_r}\!\: H\textsubscript{LC}(f)\!\: y\textsubscript{env}(f) +n\textsubscript{FS}(f)
\end{equation}
where the LC filter transfer function is $H\textsubscript{LC}(f)=1/\big[1-j(2\pi f)^2 L_f C_f\big]$ with cutoff frequency $f\textsubscript{cut}=1/\big[2 \pi \sqrt{L_f C_f}\,\big]$, and $n\textsubscript{FS}(f)$ is the Fourier transform pair of the FS PAPR estimator noise $n\textsubscript{FS}(t) \sim\mathcal{N}(0,\sigma\textsubscript{FS}^{2})$. In the above, we have assumed the first-order LC filter for mathematical simplicity, but it can easily be generalized to the other type of practical filters for circuit implementation. 

With proper selection of $L_f$ and $C_f$ values, we can filter out the DC component in (\ref{eqn:fDomainSignal}) which is then approximated to 
\begin{equation}
y\textsubscript{FS}(f) = \sqrt{1-\rho_r}\:\! \big[\!\: y\textsubscript{env}(f)-y\textsubscript{env}(0) \big] + n\textsubscript{FS}(f).
\end{equation}
The time-domain expression of $y\textsubscript{FS}(f)$ is obtained using the inverse Fourier transform as 
\begin{equation}
\begin{aligned}
y\textsubscript{FS}(t) &= \sqrt{1-\rho_r}\left[ y\textsubscript{env}(t)-\frac{1}{T}\int_{T} y\textsubscript{env}(t)\!\: dt \right] + n\textsubscript{FS}(t)\\
&= \sqrt{1-\rho_r}\!\: \big[\!\: y\textsubscript{env}(t)-\bar{y}\textsubscript{env}\!\: \big] + n\textsubscript{FS}(t)
\end{aligned}
\end{equation}
where $\bar{y}\textsubscript{env}$ indicates the DC component of the signal envelope. 
After subtracting out the DC component from $y\textsubscript{env}(t)$, $y\textsubscript{FS}(t)$ becomes bipolar signal, which is similar to AM demodulation process. Furthermore, the subtraction of the DC component incurs distortion in the PAPR value of multi-tone shaped signal at the FS PAPR estimator, as illustrated in blue dashed line in Fig.~\ref{fig:Rx}(b). This distortion results in the reduced PAPR for multi-tone shaped signal, which yields $\text{PAPR}\textsubscript{FS}< 2N$. For this reason, we consider only a large $\rho$ value (corresponding to single tone shaped signal) for ID at the FS PAPR estimator. 

Following the approach for the PS PAPR estimator, if we assume a linear HPA and ignore noise, we can express $y\textsubscript{FS}(t)$ as 
\begin{equation}
\label{eqn:FSsignal}
\begin{aligned}
y\textsubscript{FS}(t) 
&= \sqrt{1-\rho_r}\!\: |h|\!\: g\left\{ \big\lvert s_s(t) + s_m(t)\big\rvert - \frac{1}{T}\int_{T} |s(t)|\!\: dt \right\} \\
&= \sqrt{1-\rho_r}\!\: |h|\!\: g\,\bigg\{ \Big| \sqrt{2\rho P\textsubscript{dr}} + \text{Re}\{s_m(t)\} \\
&\qquad \qquad \qquad+j\!\: \text{Im}\{ s_m(t)\} \Big| - \frac{1}{T}\int_{T} |s(t)|\!\: dt \bigg\}.
\end{aligned}
\end{equation}
Since the signal envelope can be evaluated using $|s(t)|^2 = \big(\text{Re}\{s(t)\}\big)^2+\big(\text{Im}\{s(t)\}\big)^2$, (\ref{eqn:FSsignal}) can be rewritten as
\begin{equation}
\label{eqn:FSReIm}
\begin{aligned}
y\textsubscript{FS}(t) &= \sqrt{1-\rho_r}\!\: |h|\!\: g \\
&\times \bigg\{ \sqrt{\big( \sqrt{2\rho P\textsubscript{dr}} + \text{Re}\{s_m(t)\} \big)^2 +\big( \text{Im}\{s_m(t)\} \big)^2} \quad \\ 
&\qquad \qquad \qquad \qquad \qquad \qquad - \frac{1}{T}\int_{T} |s(t)|\!\: dt \bigg\}.
\end{aligned}
\end{equation}
With large $\rho$ assumption, that is similar to preventing the over-modulation in AM, if we can ignore the imaginary part of (\ref{eqn:FSReIm}), it can be simplified to  
\begin{equation}
\label{eqn:FSsim}
\begin{aligned}
y\textsubscript{FS}(t) &=  \sqrt{1-\rho_r}\:\! |h|\!\: g \\
&\times \bigg\{ \text{Re}\{s_m(t)\} + \sqrt{2\rho P\textsubscript{dr}} - \frac{1}{T}\int_{T} |s(t)|\!\: dt \bigg\}. 
\end{aligned}
\end{equation}
Note that $2\rho P\textsubscript{dr}=\frac{1}{T}\int_{T} |s_s(t)|^2 dt \approx \frac{1}{T}\int_{T} |s(t)|^2 dt$ with large $\rho$, and $\frac{1}{T}\int_{T} |s(t)|^2 dt \geq \big(\frac{1}{T}\int_{T} |s(t)|\!\: dt\big)^2$ by Cauchy-Schwarz inequality. Thus, the relationship between the envelope of single tone $\sqrt{2\rho P\textsubscript{dr}}$ and the DC component $\bar{y}\textsubscript{env}$ is given as 
\begin{equation}
\frac{1}{T}\int_{T} |s_s(t)|^2 dt \approx \frac{1}{T}\int_{T} |s(t)|^2 dt \geq \Big( \frac{1}{T}\int_{T} |s(t)|\!\: dt \Big)^2. 
\end{equation}
Here, the equality holds as $\rho \rightarrow 1$. This means that we can cancel out the last two terms in (\ref{eqn:FSsim}), which allows to decompose the multi-tone data embedded signal and the single tone power signal with large $\rho$ assumption, similar to that in \cite{KWC}. Therefore, with large $\rho = \rho\textsubscript{FS}$, we can derive 
\begin{equation}
\begin{aligned}
y\textsubscript{FS}(t) &\approx  \sqrt{1-\rho_r}\!\: |h|\!\: g\, \text{Re}\{s_m(t)\} \\
&=\sqrt{\frac{2\!\: (1-\rho_r)(1-\rho\textsubscript{FS})P\textsubscript{dr}}{N}}\, |h|\!\: g \sum_{n=1}^{N} \cos \big(2\pi  f_{n} t\big)
\end{aligned}
\end{equation}
which is a scaled version of multi-tone, and the average power is $\mathcal P\textsubscript{FS}(\rho_\text{FS},N) = (1-\rho_r)(1-\rho_\text{FS})|h|^2\mathcal P(\rho\textsubscript{FS},N) + \sigma\textsubscript{FS}^{2}$. In Fig.~\ref{fig:Rx}(c), we observe that the input signal at the FS PAPR estimator (blue solid line) is obtained by subtracting the DC component from the detected baseband envelope. 
The PAPR of $y\textsubscript{FS}(t)$ is then calculated as 
\begin{equation}
\begin{aligned}
\text{PAPR}\textsubscript{FS}(N)&=\frac{\max_{t}|y\textsubscript{FS}(t)|^2}{\mathbb{E} \big[|y\textsubscript{FS}(t)|^2\big]}\\
&= \frac{\Big(\sqrt{2\!\: (1-\rho_r)(1-\rho\textsubscript{FS})N P\textsubscript{dr}}\,\Big)^2}{(1-\rho_r)( 1-\rho\textsubscript{FS})P\textsubscript{dr}} \approx 2N. 
\end{aligned}
\end{equation}
Consequently we validate $\text{PAPR}\textsubscript{FS} = 2N$, which is equivalent to the passband PAPR value without requiring the factor of 2. This is because the FS PAPR estimator utilizes only the real part of the received signal. 

From the above receiving process at each PAPR estimator, we observe that the PS PAPR estimator is suitable for multi-tone shaped signal with $\text{PAPR}\textsubscript{PS}\leq2N$, whereas the FS PAPR estimator is for single tone shaped signal with $\text{PAPR}\textsubscript{FS}\leq2N$. Each equality holds when $\rho=0$ and $\rho=\rho\textsubscript{FS}$, respectively. 
Based on the above observation, we can choose the maximum value between the outputs of each PAPR estimator as 
\begin{equation}
\text{PAPR}\textsubscript{ID}=\max \big\{ \text{PAPR}\textsubscript{PS},\text{PAPR}\textsubscript{FS} \big\}.
\label{fpapr}
\end{equation}
Here, if we consider the HPA nonlinearity at the transmitter, the instantaneous power will likely be saturated at large $N$ when the instantaneous amplitude of the HPA input signal exceeds $A\textsubscript{sat}$. 
Hence, unlike the ideal case where the transmit PAPR is $\text{PAPR}\textsubscript{TX}(N)=2N$ as shown in \cite{DIK}, the HPA distorts the amplitude of the input signal. Thus, the transmit PAPR (and the resulting $\text{PAPR}\textsubscript{PS}$ and $\text{PAPR}\textsubscript{FS}$) are no longer linearly proportional to the number of tones. Due to this HPA nonlinearity, it is intractable to evaluate the receive PAPR in closed form. But we can still invoke a statistical approach to analyze the error-rate performance of PAPR modulation with the HPA nonlinearity. 

For analysis of the SER, conditioned on the channel gain $|h|$, the cumulative distribution function (CDF) of PAPR\textsubscript{ID} in (\ref{fpapr}) is defined as
\begin{equation}
\begin{aligned}
F\textsubscript{ID}(\gamma,N)&=\text{Pr}\Big\{\text{PAPR}\textsubscript{ID}(N)<\gamma \,\big|\, |h| \Big\}\\
&=\text{Pr} \Big\{\!\max \big\{\text{PAPR}\textsubscript{PS},\text{PAPR}\textsubscript{FS}\big\} \,\big|\, |h| \Big\}.
\end{aligned}
\end{equation}
Since $n\textsubscript{PS}(t)$ and $n\textsubscript{FS}(t)$ are statistically independent, \text{PAPR}\textsubscript{PS} and \text{PAPR}\textsubscript{FS} are also independent. 
Thus, the CDF is given by 
\begin{equation}
\begin{aligned}
F\textsubscript{ID}(\gamma,N)&=\text{Pr}\Big\{\text{PAPR}\textsubscript{PS}(N)<\gamma \,\big|\, |h| \Big\} \\
&\qquad \times \text{Pr}\Big\{\text{PAPR}\textsubscript{FS}(N)<\gamma \,\big|\, |h| \Big\} \\
&=F\textsubscript{PS}(\gamma,N) \times F\textsubscript{FS}(\gamma,N)
\end{aligned}
\label{eqn:CDFanal}
\end{equation}
where $F\textsubscript{PS}(\gamma,N)$ and $F\textsubscript{FS}(\gamma,N)$ represent the CDF's of each PAPR estimator, respectively.

Based on this, the CDF of PAPR modulation at each PAPR estimator can be derived as 
\begin{equation}
\label{eqn:CDF_PAPR}
\begin{aligned}
F_i(\gamma,N)=\prod_{t} \Bigg\{1-\int_{0}^{\infty} f_{|h|}(z) \times Q_{1/2} \Big(\!\sqrt{\lambda_i},\sqrt{\nu_i}\,\Big)\!\: dz \Bigg\}
\end{aligned}
\end{equation}
where $i\in\{\text{PS},\text{FS}\}$ for each estimator, and $f_{|h|}(z)$ denotes the probability density function (PDF) of channel fading. Also, we define $Y\textsubscript{PS}(t)=\sqrt{\rho_r}\, y\textsubscript{env}(t)$ and $ Y \textsubscript{FS}(t)=\sqrt{1-\rho_r}\!\:\big[ y\textsubscript{env}(t)-\bar{y}\textsubscript{env}\big]$. Then $\nu_i=\gamma\, \mathcal P_i(\rho,N)/\sigma_{i}^2$ and $\lambda_i= \big[ Y_i(t)/\sigma_{i} \big]^2$. 
For the detailed proof, please see Appendix \ref{apd:CDF_PAPR} with \cite{JJPlet, IK}. 

Using the above CDF, the SER of the PAPR modulation with total $Q$ tones is evaluated as
\begin{equation}
P\textsubscript{SER}(\rho, Q)=\frac{1}{Q} \sum_{q=1}^Q p(q)
\end{equation}
where $p(1)=1-F\textsubscript{ID}(3,1)$, $p(Q)=F\textsubscript{ID}(2Q-1,Q)$, and $p(N)=1-F\textsubscript{ID}(2N+1,N)+F\textsubscript{ID}(2N-1,N)$ for $1<N<Q$.

\subsection{Nonlinear Energy Harvesting}
\label{sec:EH}
\begin{figure}
\centering
\subfigure[]{
\includegraphics[width=0.5\textwidth]{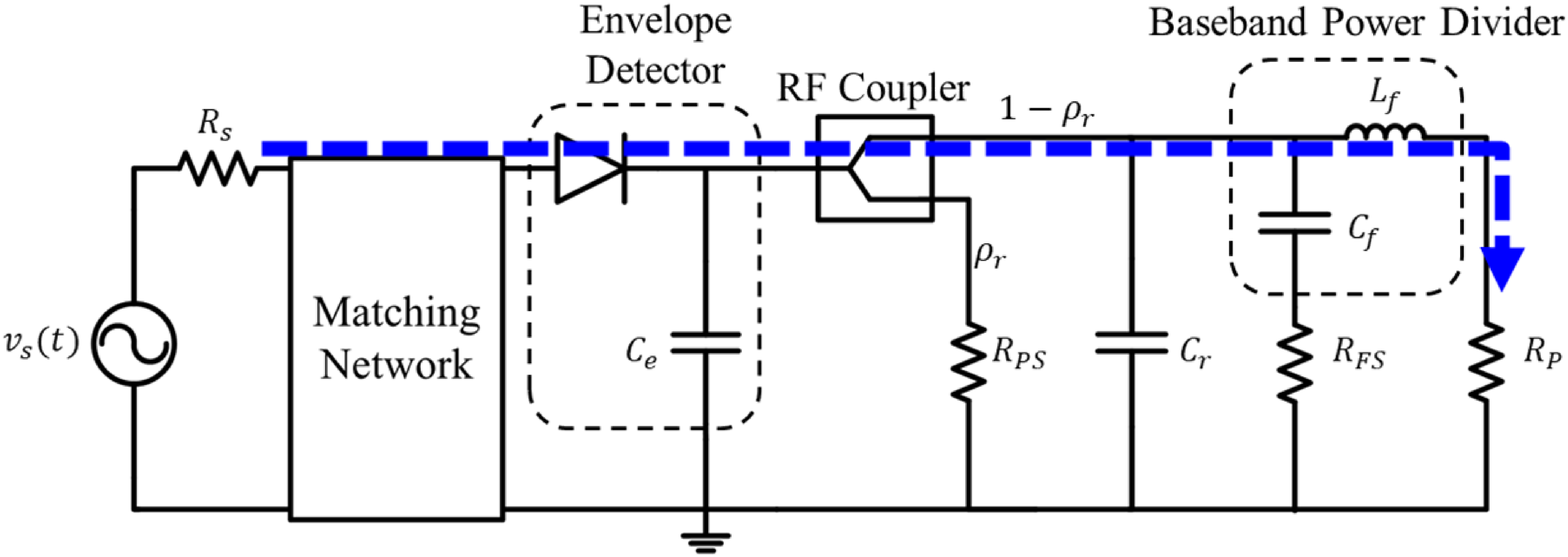}
}
\subfigure[
\hspace{-0.15in}]{
\includegraphics[width=2.5in]{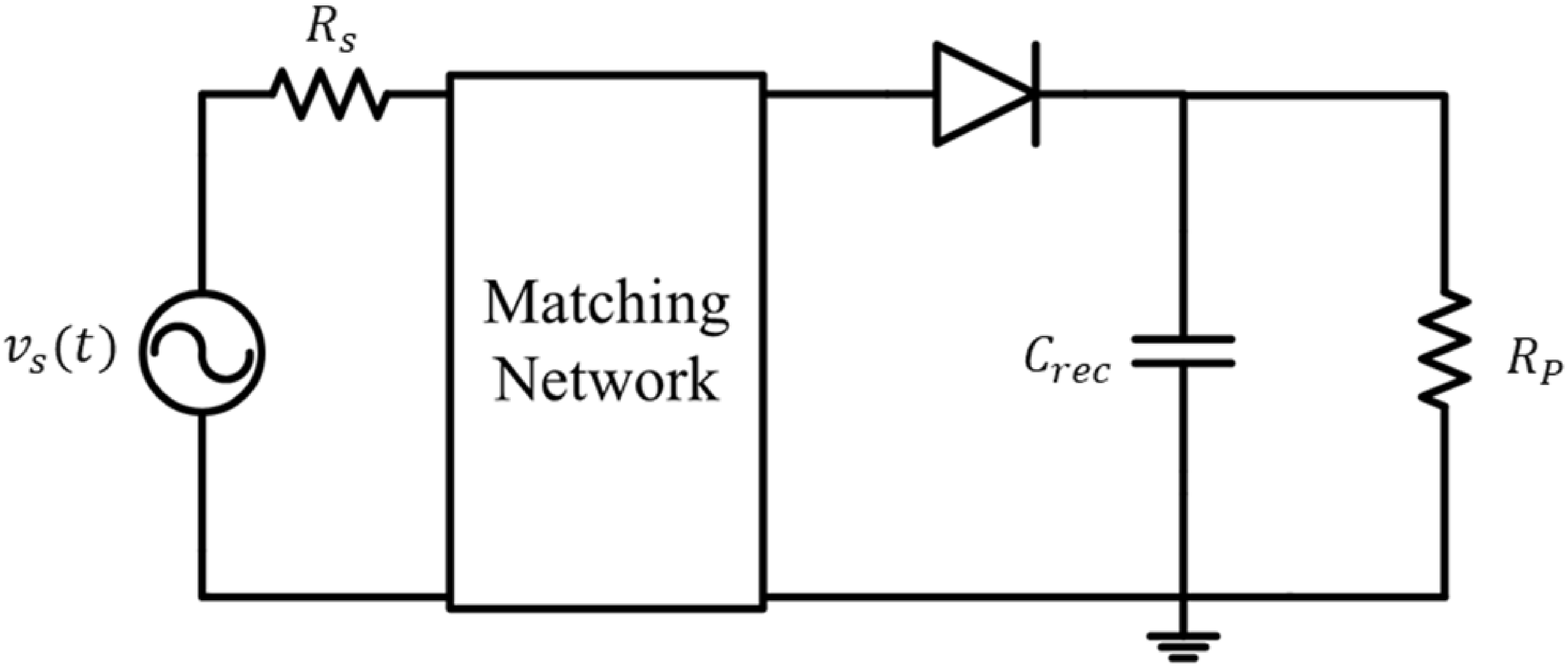}
}
\subfigure[]{
\includegraphics[width=3.6in]{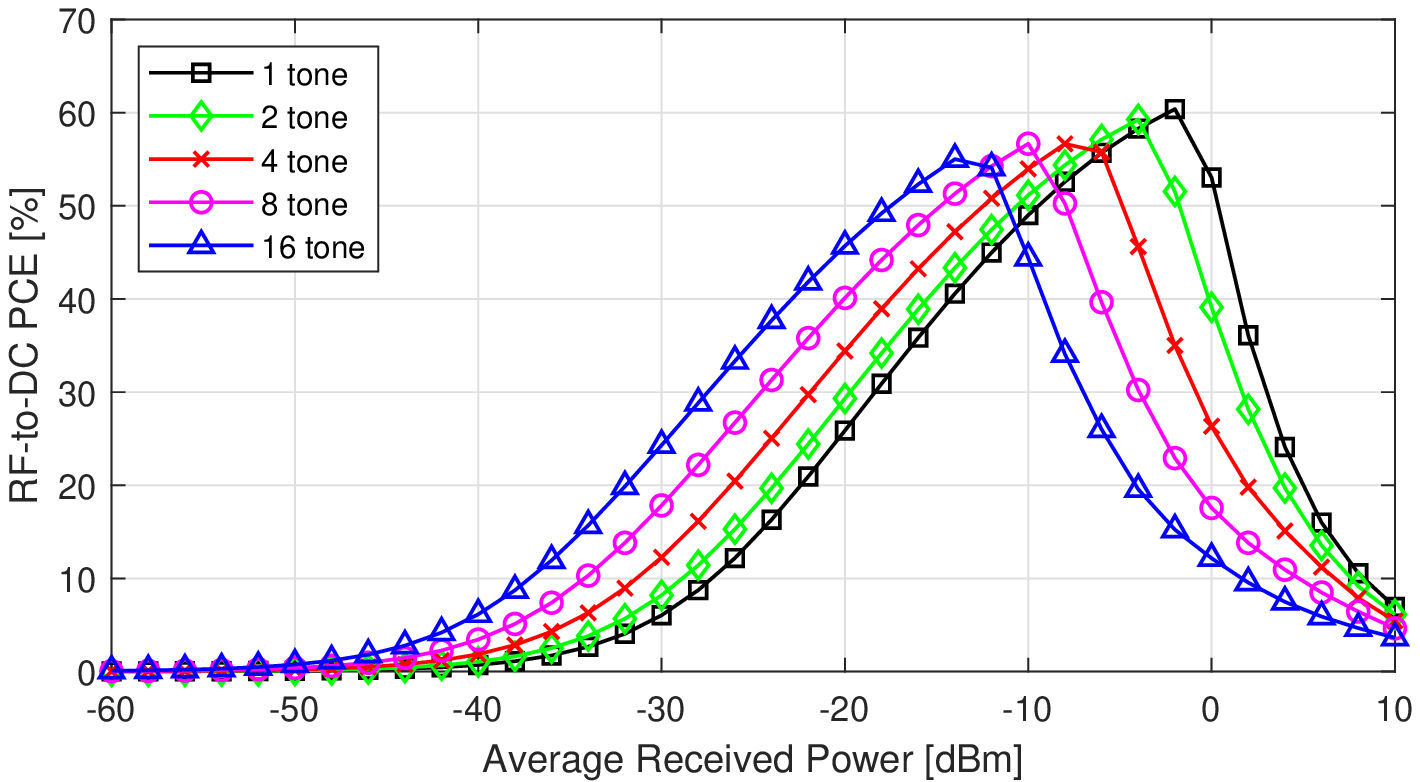}
}
\caption{Circuit of (a) proposed unified receiver and (b) equivalent rectifier circuit  with matching network. (c) RF-to-DC PCE curve based on the piecewise linear EH model ($Q=1, 2, 4, 8, 16$). Data was obtained by the ADS circuit simulation.}
\label{fig:circuit}
\end{figure}

The unified SWIPT receiver circuit is implemented in Fig.~\ref{fig:circuit}(a), where the antenna circuit is commonly modeled by Thevenin equivalent voltage source $v_s(t)$ in series with an impedance $R_s$. 
The maximum power is then delivered from the antenna to the envelope detector when the impedance is perfectly matched. The envelope detector tracks the signal envelope with capacitor $C_e$, which converts RF signal to the baseband output signal. After the envelope detector, the signal power is split into $\rho_r$ by the RF coupler which is a static power splitter. Since $\rho_r$ is small, the most of the signal power is fed to a smoothing capacitor $C_r$ to shape the signal while the rest of the signal is used for PS PAPR estimation at $R\textsubscript{PS}$. The shaped signal after the smoothing capacitor is divided into two branches, in a way similar to the FS architecture in \cite{KWC}, one for the FS PAPR estimator load $R\textsubscript{FS}$ and the other for the energy harvester load $R_P$. If we apply the DC analysis for wireless power transfer (WPT) which is the blue dashed arrow in Fig.~\ref{fig:circuit}(a), the receiver circuit is reduced to the equivalent rectifier circuit in Fig.~\ref{fig:circuit}(b) with equivalent capacitance $C\textsubscript{rec}=C_e+C_r+C_f$.

Note that we have assumed a linear operating region of the diode envelope detector for ID at the unified receiver in Section \ref{sec:UnifiedReceiver}. But the unified receiver should always harvest energy regardless of the input power range. In other words, the unified SWIPT receiver operates EH mode even when the received power is too low or high (i.e., square-law or saturation region). Thus, we have to consider the nonlinearity of the diode rectifier (envelope detector) over wide input power range for EH at the unified receiver.

To characterize the diode nonlinear behavior (e.g., turn-on sensitivity and saturation), \cite{BCwpt} and \cite{EB} proposed the nonlinear models based on Taylor series approximation of diode small signal equation and logistic curve fitting for measured experimental data, respectively. However, those models may not be accurate when both saturation effect and small signal model are considered for wide input power. Also, to maximize the efficiency of the rectifier, various circuit designs (e.g., reconfigurable rectifier, deep learning based rectifier, etc.) are proposed in \cite{PX,WWYL,DK,MAA}, which are far from the simple diode rectifier. Thus, the diode based mathematical model has limitation to precisely predict the state-of-the-art results of practical EH circuit design. 

We adopt a \textit{piecewise linear EH model} in \cite{PNA} which approximates the nonlinear energy harvester based on circuit measurement or simulation data to address the nonlinear characteristics of various rectifier designs as well as simple diode rectifier. Here, the EH model reflects two important nonlinear features such as turn-on sensitivity and saturation. If the input power is less than the turn-on sensitivity $P_\text{on}$, the EH circuit cannot be activated. Meanwhile, the harvested power becomes saturated with the maximum harvested power $P^\text{max}_\text{EH}$ if the input power exceeds $P_\text{sat}$. 
If we let input $x=P_\text{in}$, the harvested power that results from using $q$-tone multisine waveforms can be expressed as 
\begin{equation}
P^q_\text{EH}(x)=
\begin{cases}
0 & x \in [0,x^q_{0}),\\
\eta^q_k(x-x^q_{k-1})+y^q_{k-1} & x \in [x^q_{k-1},x^q_{k}),\\
y^q_K & x \in [x^q_{K},\infty)
\end{cases}
\end{equation}
where $\{x^q_k\}^{k=K}_{k=0}$ and $\{y^q_k\}^{k=K}_{k=0}$ denote the sets of supporting points with $x^q_0=P^{q}_\text{on}$, $x^q_K=P^{q}_\text{sat}$,  and $y^q_K=P^{\text{max},q}_\text{EH}$.

Fig.~\ref{fig:circuit}(c) shows the PCE curve of the piecewise linear EH model with single tone/multi-tone waveforms ($Q=1, 2, 4, 8 ,16$). The data point of the harvested power is obtained through ADS circuit simulation. For the rectifier, we have used the Skyworks SMS7630 Schottky diode rectifier to simplify the receiver circuit.\footnote{However, if other complex rectifier design is utilized to enhance the EH efficiency, we can still adopt this EH model by fitting the model to data from specsheet or measurement, because the piecewise linear EH model is based on measurement of the EH circuit, not the analysis of circuit operation.} 
A matching network is tuned at $-10$dBm input power with series capacitor $C_m=0.25$pF and shunt inductor $L_m=11.67$nH. Also, the equivalent capacitor is set to $C_\text{rec}=1$nF and the energy harvester load $R_P=1$k$\Omega$. Note that we have converted the harvested power versus input power relationship to the RF-to-DC PCE versus input power one, defined by $\eta=P_\text{EH}/P_\text{in}$. 
Since the harvested power is normalized by the input power, it shows accurate EH efficiency regardless of the input power range. It can be observed that multi-tone shows better PCE in low input power region. On the other hand, single tone shows better PCE due to the saturation of the diode when the input power is high.

\section{Temporal Convolutional Network Based Adaptive Control Algorithm}
\label{sec:TCNalgorithm}

\begin{figure*} 
\centering
\includegraphics[width=0.8\textwidth]{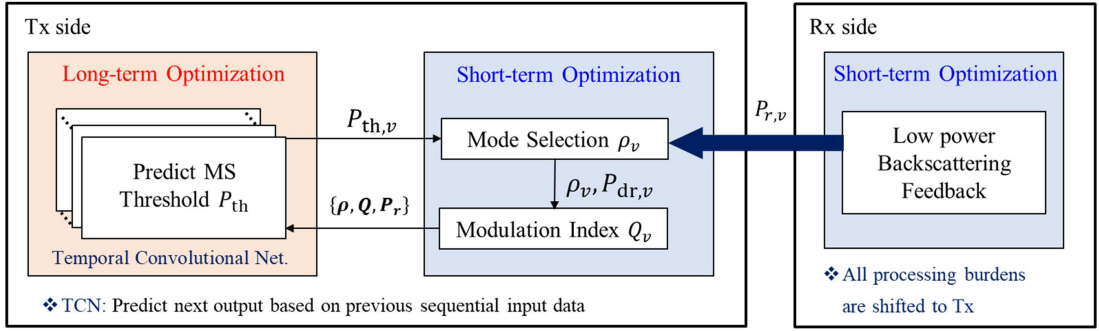}
\caption{Functional description of the proposed temporal convolutional network (TCN) based adaptive control algorithm for the unified SWIPT.}
\label{fig:flow}
\end{figure*}

Because of the HPA and rectifier nonlinearity, there exists the rate-energy tradeoff between the single tone and multi-tone modes of the proposed unified SWIPT. The former can achieve a higher data rate compared to the latter, but the RF-to-DC PCE is not optimal at low received power due to the nonlinear characteristics of the rectifier. Unlike this, the latter is suitable for low received power with enhanced RF-to-DC PCE of the rectifier but more susceptible to nonlinear distortion of the HPA, resulting in a lower data rate at the high HPA drive power. Therefore, it is crucial to adjust the communication mode for optimizing performance, considering the end-to-end efficiency of the unified SWIPT system.

For switching the communication mode adaptively, we define the MS threshold $P\textsubscript{th}$ so as to achieve the maximum data rate. For instance, when the received power feedback $P_{r,v}$ is greater than the MS threshold ($P_{r,v} \geq P\textsubscript{th}$), the single tone signal is utilized ($\rho =\rho\textsubscript{FS}$) and otherwise the multi-tone signal ($\rho=0$). It is obvious that both the harvested power and achievable rate are affected largely by updating the MS threshold in a time window. Based on the modulation index (the number of tones) $Q$ and power allocation ratio $\rho$ which are selected by adaptive control with $P\textsubscript{th}$, the achievable rate of the unified SWIPT with outage probability is evaluated as
\begin{equation}
R_{v} =\frac{1}{T}\big[1-p\textsubscript{out}(\rho, Q)\big]\log_2Q.
\end{equation}
Note that the outage probability $p\textsubscript{out}(\rho, Q)$ is defined as $p\textsubscript{out}(\rho, Q)=\Pr[P\textsubscript{SER}(\rho, Q) > \text{SER}\textsubscript{tag}]$ for a given $\rho$ and $Q$, which is the SER exceeding the target SER. 

An optimization problem for adaptive control to maximize the average achievable rate with energy constraint is given as 
\begin{equation*}
\begin{aligned}
\text{(P1)} :~
& \max_{P_{\text{th},v}}
&& \!\!\! \mathbb{E}_{v} [R_{v}] \\
& \text{ s.t.}
&& \!\!\! P\textsubscript{EH}\geq P_{C}\\
\end{aligned}
\end{equation*}
 where $P_C$ represents the circuit power consumption at the receiver. Note that (P1) can be viewed as mixed-time scale optimization which is non-convex and difficult to solve directly as both objective and constraint functions are combinatorial with discrete MS attributes $\rho$ and $Q$. Also, the objective function involves the expectation over the channel index $v$. On the other hand, the MS attributes can vary over every channel block, as they are greatly affected by the end-to-end efficiency (i.e., RF-to-DC PCE of the rectifier and the HPA nonlinear distortion of each mode). 
  
\begin{figure*} 
\centering
\includegraphics[width=0.8\textwidth]{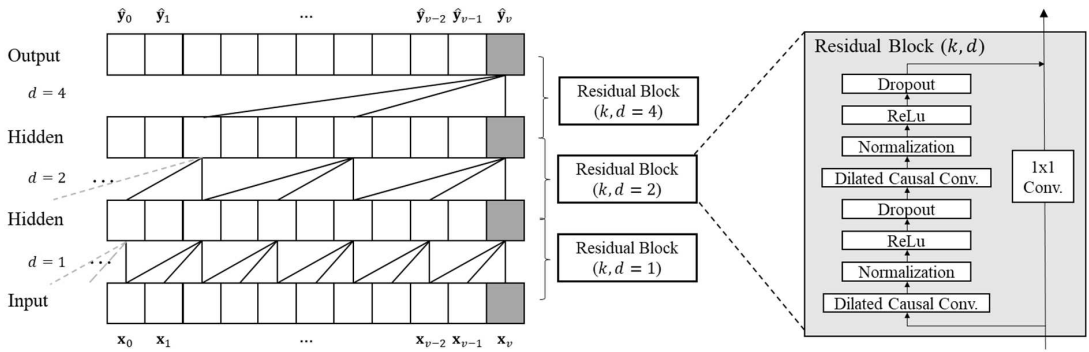}
\caption{Illustration of TCN architecture and example of residual block with dilated causal convolution, where the dilation factor is $d=[1, 2, 4]$ with filter size $k=3$. For residual connection, the 1$\times$1 convolution is added when input and output have different dimensions.}
\label{fig:TCN}
\end{figure*}

Moreover, in practice, the channel exhibits some degree of temporal correlations because of the mobility of devices and time-varying wireless environments. To model the temporal correlation between fading blocks, the Gauss-Markov channel model is considered as $h_v=\zeta h_{v-1}+u_v$, where $\zeta\in [0,1]$ is the temporal correlation coefficient and $\{u_v\}$ are the CSCG random variables with zero mean and variance $(1-\zeta^2)$. Note that the channel is time-varying with respect to the channel block index $v$, which affects the received power feedback $P_{r,v}$. Therefore, it is necessary to switch the communication mode adaptively to the received power feedback which reflects the temporal correlation of the time-varying channel. Here, the receiver feedbacks the measurement using the monostatic backscatter prior to each frame transmission. 
To minimize the HPA distortion to the pilot signal for backscatter and maintain a constant transmit power level for feedback, the transmitter always sends the pilot signal with the reference power $P\textsubscript{ref}$, that is, a back-off output power level of the HPA below the saturation point at which the amplifier will continue to operate in the linear region. Thus, the received power feedback at the $v$th fading block is evaluated as $P_{r,v}=|h_v|^2 P\textsubscript{ref}$.

To solve the optimization problem (P1), we divide (P1) into iterative short-term/long-term optimization. Here, $P\textsubscript{th}$ is updated on a long-term scale using a deep learning technique while MS attributes ($\rho$, $Q$) are determined on a short-term scale, as shown in Fig.~\ref{fig:flow}. Especially, to handle both the HPA and rectifier nonlinearity, and also adaptively controlling the communication mode depending on the temporal channel variation, we adopt the TCN based asymmetric processing. For instance, the short-term optimization at the $v$th fading block selects the power allocation ratio $\rho_v$ by comparing $P_{r,v}$ with $P_{\text{th},v}$, and the modulation index $Q_v$ is set by maximizing the data rate while satisfying the target SER. In long-term optimization, the previous estimated MS threshold $\boldsymbol{\hat P_\bold{th}}$, the short-term optimization results (i.e., $\boldsymbol{\rho}$, $\boldsymbol{Q}$), and the received power measurements $\boldsymbol{P_r}$ are all used to update the MS threshold by the TCN. Here, we take the input data to the TCN with sequence length $W$, namely the previous $W$ MS threshold estimates $\boldsymbol{\hat P_\bold{th}}=\{\hat P_{\text{th},v-w}\}_{w=1}^{W}$ and the previous $W$ MS attributes with the received power measurement $\{\boldsymbol{\rho},\boldsymbol{Q},\boldsymbol{P_r}\}=\{\rho_{v-w}, Q_{v-w}, P_{r,v-w}\}_{w=1}^{W}$, where $w\in[v-W,v-1]$ is a sliding window over $v$. Note that all processing burden of the proposed algorithm is shifted to the transmitter, so as to lower the receiver complexity, by which the transmitter optimizes the modulation index and power allocation ratio based on the received power feedback, leading to {\it transmitter-oriented asymmetric processing}. 

We first look into the short-term optimization for a given MS threshold $P_{\text{th},v}$ which is updated by the long-term optimization at the $v$th channel block. $\rho_v$ and $Q_v$ are selected by the short-term optimization problem, which is formulated as 
\begin{equation*}
\begin{aligned}
\text{(P2)} :~
&\max_{Q_v} && \!\!\! R_{v} \\
& \text{ s.t.}
&& \!\!\! \rho_v=\rho\textsubscript{FS}I_{P_{\text{th},v}}(P_{r,v})\\
&&&\!\!\! P\textsubscript{EH}\geq P_{C}. 
\end{aligned}
\end{equation*}
In the above, the inequality constraint is the energy-causality condition for self-powering at each fading block, and $I_{P_{\text{th},v}}(x)$ is the indicator function such that $I_{P_{\text{th},v}}(x)=1$ for $x \in \{x|x\geq P_{\text{th},v}\}$ and otherwise $I_{P_{\text{th},v}}(x)=0$. Then, if the received power is greater than the MS threshold ($P_{r,v} \geq P_{\text{th},v}$), the single tone mode ($\rho_v=\rho\textsubscript{FS}$) is selected and otherwise the multi-tone mode ($\rho_v=0$). The objective function of (P2) is monotonic increasing with the received power and $Q_v$. Once the received power measurement $P_{r,v}$ is given at $v$th block, the above optimization problem is the integer programming as the modulation index $Q_v$ is discrete and a finite set. Thus, we can solve (P2) using simple integer programming optimization techniques such as branch-and-bound algorithm.

Next, the TCN based long-term optimization updates the MS threshold $P_{\text{th},v}$ for iterative operation. The TCN has an architecture that the previous sequential input data affects the prediction of the next output. In general, the sequence modeling problem with input sequence data $\{ \bold x_1, ... ,\bold x_W\}$ for a sequence length $W$ is a function $\mathcal F$ that produces the mapping
\begin{equation}
\bold {\hat{y}}_1, ... ,\bold{\hat{y}}_W = \mathcal F(\bold x_1, ... ,\bold x_W;\theta_{\mathcal F}),
\label{eqn:mapping}
\end{equation}
which minimizes the loss function between the actual outputs $\{\bold {y}_v\}$ and the predictions $\{\bold {\hat{y}}_v\}$. 
For the sequence modeling of the TCN, we define $\mathcal F(\cdot;\theta_{\mathcal F})$ as the overall mathematical functions of the TCN model, in which $\theta_{\mathcal F}$ is the set of the parameters (i.e., weights and biases) of the corresponding functions. We utilize the mean-square error (MSE) as the loss function which can also represent the deep learning performance. The MSE loss function is defined as
\begin{equation}
\begin{aligned}
\label{eqn:loss_ft}
\mathcal L(\theta_\mathcal F)&=\frac{1}{|\mathcal T|} \sum_{v\in \mathcal T} \big{\|} \!\:\bold {y}_v-\bold {\hat{y}}_v \big{\|}^2 \\
&=\frac{1}{|\mathcal T|} \sum_{v\in \mathcal T} \big{\|} \!\:\bold {y}_v-\mathcal F(\{\bold x_{v-w+1}\}_{w=1}^{W};\theta_{\mathcal F}) \big{\|}^2
\end{aligned}
\end{equation}
where $\mathcal T$ denotes the set of training samples. By minimizing the loss function of (\ref{eqn:loss_ft}) with respect to $\theta_\mathcal F$, the proposed algorithm can be optimized iteratively using the stochastic gradient descent method.

Based on the above framework, we employ the TCN for the proposed adaptive control algorithm, whose architecture is illustrated in Fig.~\ref{fig:TCN}. The TCN uses 1-D fully convolutional network (FCN) architecture, where each hidden layer is the same as the input layer. Thus, the FCN produces an output with the same length as the input similarly to conventional RNN model such as LSTM. Note that the 1-D convolution is suitable for real-time applications due to the low computational requirement of array operation (not the matrix operation of CNN).  Also, the TCN uses causal convolution to prevent the leakage of information from the future to the past. There are no previous data missed or future data produced when the information is transferred between the layers of the convolutional network. In general, the causal convolution can only look back at a certain history size which is proportional to the depth of the network. Thus, an extremely large size of deep network or convolution filter is required to achieve a long effective history size or memory, which significantly increases the computational complexity. 

To tackle this problem, we employ dilated convolution \cite{AVDO} which can effectively capture the long-term patterns of the input data by enabling an exponential receptive field. More formally, for the sequence input $\bold x \in \mathbb R^n$ and the convolution filter $f:\{0,...,k-1\} \rightarrow \mathbb R$ with size $k$, the dilated convolution operation $D(\cdot)$ on the sequence element $s$ can be defined as
\begin{equation}
D(s) = \sum_{\kappa=0}^{k-1} f(\kappa) \: \bold x_{s-d\times\kappa}
\end{equation}
where $d$ is the dilation factor exponentially increasing with the depth of the network (i.e., $d=2^l$ at layer $l$ of the network). For example, the dilation factor is $d=[1, 2, 4,...,2^{L-1}]$ for the network with $L$ layers. Dilation consists of skipping $d$ values between the inputs of the convolutional operation, as shown in Fig.~\ref{fig:TCN}. Note that the dilated convolution reduces to a regular convolution when $d=1$.

To increase the receptive field of the TCN, we should choose a larger filter size $k$ or increase the depth of the network with a large dilation factor. However, this leads to deeper network architecture with more training parameters. For this reason, we can replace the convolutional layer with a residual block to achieve network stabilization, as shown in Fig.~\ref{fig:TCN}. A residual block adds the input $\bold x$ of the block to a series of transformation functions $\mathcal G(\cdot)$ to evaluate its output, namely residual connection, which is expressed as
\begin{equation}
\text{output} = \psi(\bold x + \mathcal G(\bold x))
\end{equation}
where $\psi(\cdot)$ denotes the activation function. More details about the residual block can be found in \cite{SB,YC}.

Note that we can utilize the TCN to solve the long-term optimization problem since the received power measurements $\boldsymbol{P_r}$, the short-term optimization results (i.e., $\boldsymbol{\rho}$, $\boldsymbol{Q}$), and the previous MS threshold $\boldsymbol{\hat P_\bold{th}}$ are all sequential data with respect to the temporal channel variations. Therefore, the input sequence data for the proposed adaptive control algorithm is $\bold{x}_w= \{\rho_w, Q_w, P_{r,w}, P_{\text{th},w} \}$. Since the transmitter performs the training of the TCN via the asymmetric processing, these input data can easily be utilized for the long-term optimization with backscatter-based channel training. Hence, the TCN is suitable for updating the MS threshold with the short-term optimization results of (P2), in conjunction with the received power feedback from low-power IoT devices.

\section{Results}
In the simulations, we compare the performance of the proposed unified SWIPT with existing SWIPT systems. For this, the bandwidth is assumed to be $B=$ 200kHz with carrier frequency $f_c =$ 2.4GHz. The antenna noise power is assumed to be $\sigma_{a}^2=-110$dBm, and the estimator noise power is $\sigma\textsubscript{PS}^2=\sigma\textsubscript{FS}^2=-100$dBm for fair comparison between PS and FS paths. Also, the tone spacing for multi-tone is assumed to be $\Delta f = 10$kHz. The modulation index of the unified SWIPT is set to $Q=$ 4, 8, 16. Thus, we use 16 tones among the 20 available tones, and the edge of the bandwidth is unused, which can be regarded as guard band. For the HPA nonlinearity, the parameters are assumed to be $g^2=25$ dB, $A\textsubscript{sat}^2=10$ dBm, and $\beta=2$, which are the circuit characteristics of Mini Circuits ZHL-5W-422+ HPA, and $P\textsubscript{ref}$ is 29dBm, which is $-$6dB back-off from 1dB gain compression point (P1dB) of the HPA. The distance of the transceiver is 3m with antenna gain 5dBi. For EH, we have assumed the same rectifier circuit parameter of Section \ref{sec:EH}. Also, the power-splitting ratio at the receiver is $\rho_r=10^{-3}$.

\begin{figure}
\centering
\subfigure[]{
\includegraphics[width=0.5\textwidth]{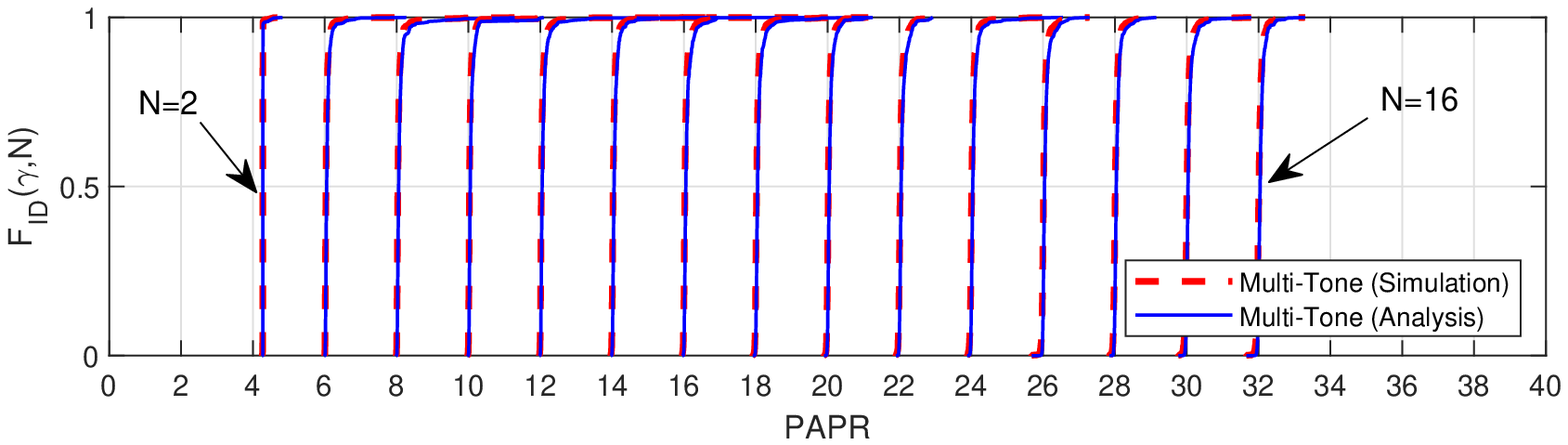}
}
\subfigure[]{
\includegraphics[width=0.5\textwidth]{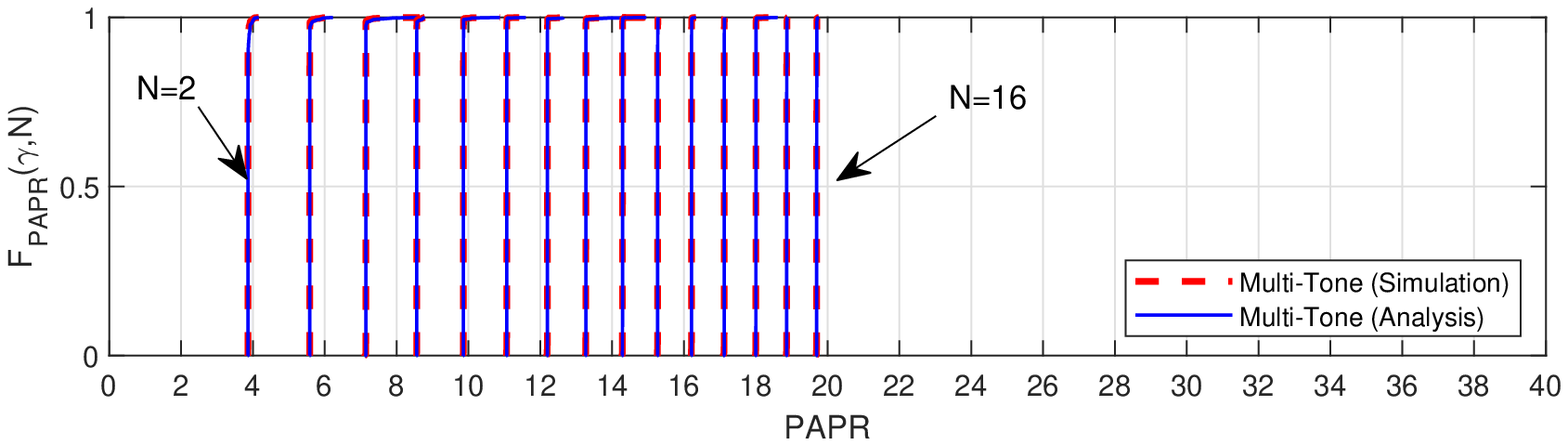}
}
\caption{CDF of the PS PAPR estimator with HPA impairment, where $P\textsubscript{dr}=-$10dBm for (a), and 0dBm for (b), respectively.}
\label{fig:CDFm}
\end{figure}
\begin{figure}
\centering
\subfigure[]{
\includegraphics[width=0.5\textwidth]{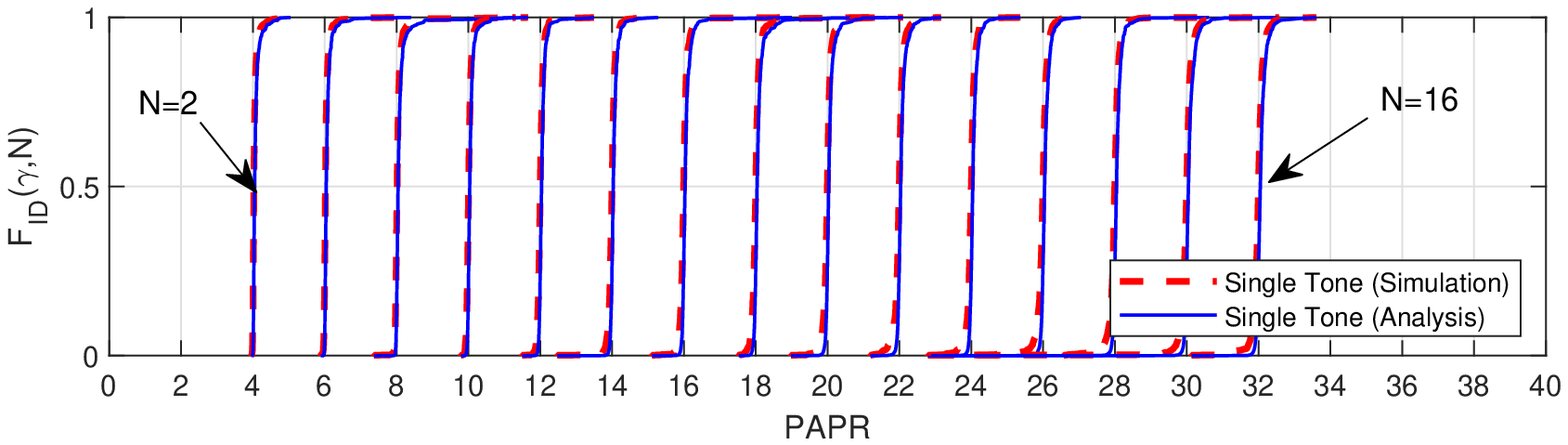}
}
\subfigure[]{
\includegraphics[width=0.5\textwidth]{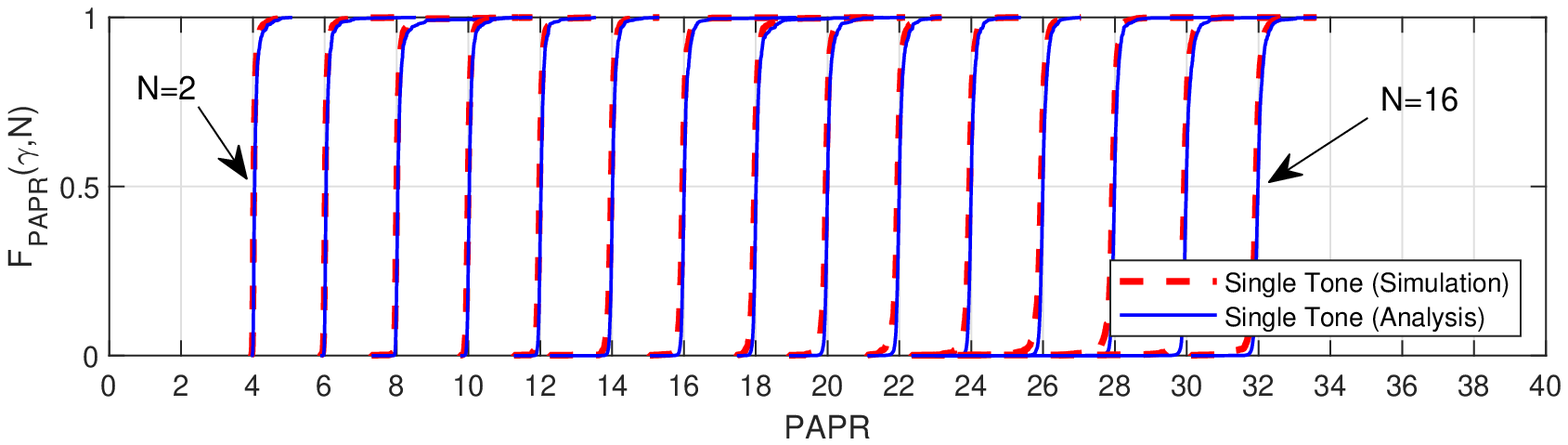}
}
\caption{CDF of the FS PAPR estimator with HPA impairment, where $P\textsubscript{dr}=-$10dBm for (a), and 0dBm for (b), respectively.}
\label{fig:CDFs}
\end{figure}

For the adaptive control, we set $\zeta =$0.99 and 0.9 for Gauss-Markov Rayleigh fading channel model with path-loss exponent 2.5.\footnote{$\zeta$ is determined by $\zeta^{T_cB}=\chi$ for the level of decorrelation $\chi$ \cite{IA}. We select $\zeta=0.9$ as the lower bound on the channel correlation because of typically $\zeta \geq 0.99$ at 2.4GHz\cite{JJP}.} Also, the convolution filter size is $k=2$, and the dilation factor of each layer is set to $d=[1, 2, 4, 8]$ for the TCN. Meanwhile, for comparison, the number of the LSTM RNN hidden layers is set to $H=4$. For both networks, the sequence length is set to the window size $W=20$. Thus, both networks have similar sequence modeling capabilities. The TCN and LSTM models are trained by using $4 \times 10^4$ training samples with 15 training epochs. The performance of the proposed algorithm is evaluated using $3.6 \times 10^5$ test samples, independent of the training samples. The outage probability $p\textsubscript{out}(\rho, Q)$ of each mode with modulation index $Q$ is evaluated through the Monte-Carlo simulations subject to the target SER $\text{SER}\textsubscript{tag}=0.01$. Finally, the circuit power consumption at the unified SWIPT receiver is set to $P_{C}=10\mu$W, which is the envelope detection based low-power IoT receiver \cite{TS}. 

Figs.~\ref{fig:CDFm} and \ref{fig:CDFs} show the CDF of the PS and FS PAPR estimators when $Q=16$. Here we have applied multi-tone shaped signal ($\rho=0$) for the PS PAPR estimator, and single tone shaped signal ($\rho=\rho\textsubscript{FS}$) for the FS PAPR estimator to assure the signal detection of each estimator. We set $\rho\textsubscript{FS}=(1-10^{-3})$ to normalize the input signal power of each PAPR estimator for fair comparison. We see that the analytical results based on (\ref{eqn:CDFanal}) well coincide with the simulation ones for all $N$. We also confirm that the estimated PAPR value at both PAPR estimators is linearly proportional to $N$ when the HPA nonlinearity is small, as shown in Figs.~\ref{fig:CDFm}(a) and \ref{fig:CDFs}(a). However, we observe that $\text{PAPR}\textsubscript{PS}$ of the multi-tone shaped signal is critically distorted when the drive power of the HPA is large (i.e., $P\textsubscript{dr}=0$ dBm), as shown in Fig.~\ref{fig:CDFm}(b). On the other hand, $\text{PAPR}\textsubscript{FS}$ of the single tone shaped signal is still linearly proportional to $N$ with FS PAPR estimator in Fig.~\ref{fig:CDFs}(b). This is because PAPR of the transmitted passband single tone signal is small due to the CW power signal, which is more tolerant to the HPA nonlinearity. For this reason, the FS PAPR estimator and single-tone shaped signal are suitable for not only conveying high power for EH but also mitigating the HPA nonlinearity on ID.

\begin{figure} 
\centering
\includegraphics[width=0.5\textwidth]{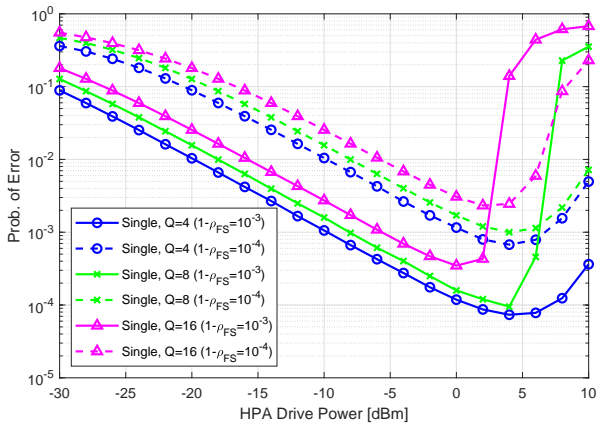}
\caption{SER performance of single tone mode for the proposed unified SWIPT with various $Q$ and $\rho\textsubscript{FS}$. }
\label{fig:SERs}
\end{figure}

In Fig.~\ref{fig:SERs}, considering the HPA nonlinearity, the single tone mode SER performance of the unified SWIPT is evaluated, which varies with the duty ratio $\rho\textsubscript{FS}$ when $Q=4, 8, 16$. We can observe that the performance is degraded as the drive power of the HPA increases because of the distortion in $\text{PAPR}\textsubscript{FS}$. We also notice that using larger $\rho\textsubscript{FS}$ results in degraded SER performance in linear operating region of the HPA because of decreased power allocation on ID. However, the distortion of the HPA is also mitigated with large $\rho\textsubscript{FS}$. This is because the decrease in transmit PAPR can reduce the voltage swing of the signal and prevent the compression or clipping of the signal waveform. Hence, we used $\rho\textsubscript{FS}=(1-10^{-4})$ for single tone mode, provided the adaptive control algorithm devised herein aims to mitigate the nonlinear distortion of the HPA.

\begin{figure} 
\centering
\includegraphics[width=0.5\textwidth]{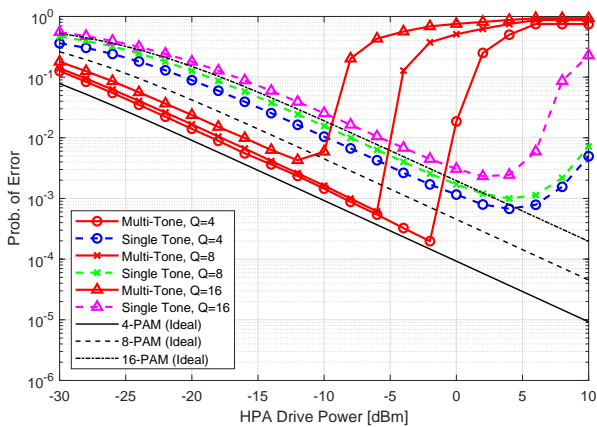}
\caption{SER performance of single tone/multi-tone modes for the proposed unified SWIPT with various $Q$.}
\label{fig:SERall}
\end{figure}

In Fig.~\ref{fig:SERall}, considering the HPA nonlinearity, the SER performance of the proposed unified SWIPT is evaluated when $Q=4, 8, 16$. The SER performance of the conventional PAM based PS SWIPT with ideal power amplifier is also plotted for comparison.\footnote{Note that the conventional data modulation schemes such as OFDM with QAM or PSK requires the {\em I/Q} based detection and FFT processing, which consume high circuit power for ID (typically from several mW to several hundreds of mW \cite{TS,TC}). So it is hard to assure the self-powering of low-power IoT devices with such modulation due to high-power consumption.} Here, the power-splitting ratio for the PAM based PS SWIPT is set to $\rho\textsubscript{PAM}=\rho\textsubscript{FS}$. We observe that the performance is degraded as $Q$ increases for both single tone and multi-tone modes because of the decreased distance between symbol (PAPR) constellation points. 
Further, these results clearly show that the SER performance of the unified SWIPT is comparable to the conventional SWIPT in linear operating range of the HPA. 

Since the maximum transmit PAPR increases as the number of tones increases, the multi-tone signal with large $Q$ is more easily distorted by the HPA. Note that the waveform of the multi-tone mode signal is clipped when the HPA is completely saturated, resulting in highly compressed transmit PAPR. In this situation, the information transmission is infeasible, which is clearly indicated by a rapid increase of the multi-tone mode SER in high drive power region of Fig.~\ref{fig:SERall}. On the other hand, we confirm that the single tone mode with FS PAPR estimator effectively mitigates the nonlinearity of the HPA, compared to the PS PAPR estimator, with slightly degraded SER performance, but it is still comparable to the conventional PAM based PS SWIPT scheme.

\begin{figure} 
\centering
\includegraphics[width=0.5\textwidth]{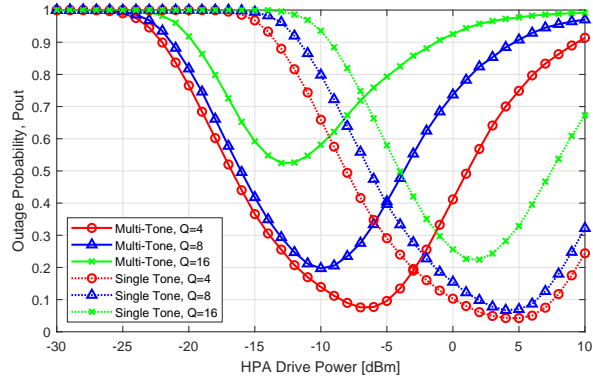}
\caption{Outage probability of single tone/multi-tone modes for the proposed unified SWIPT. The target SER is assumed to be 0.01.}
\label{fig:Pout}
\end{figure}

Based on the SER performance of single tone and multi-tone modes, the outage probability of the unified SWIPT is shown in Fig.~\ref{fig:Pout}. The results show that the multi-tone mode is suitable for low to moderate HPA drive power while the single tone mode with large power allocation ratio $\rho$ is suitable for high HPA drive power which can cause signal distortion. This tendency is consistent with the results in Fig.~\ref{fig:SERall}. We also observe that using large $Q$ to increase the data rate results in more outage for ID, which reduces the reliability of data communication. Therefore, it is important to adaptively control $\rho$ and $Q$ to optimize both the performance and reliability under the QoS of low-power IoT devices.

\begin{figure} 
\centering
\subfigure[]{
\includegraphics[width=3.6in]{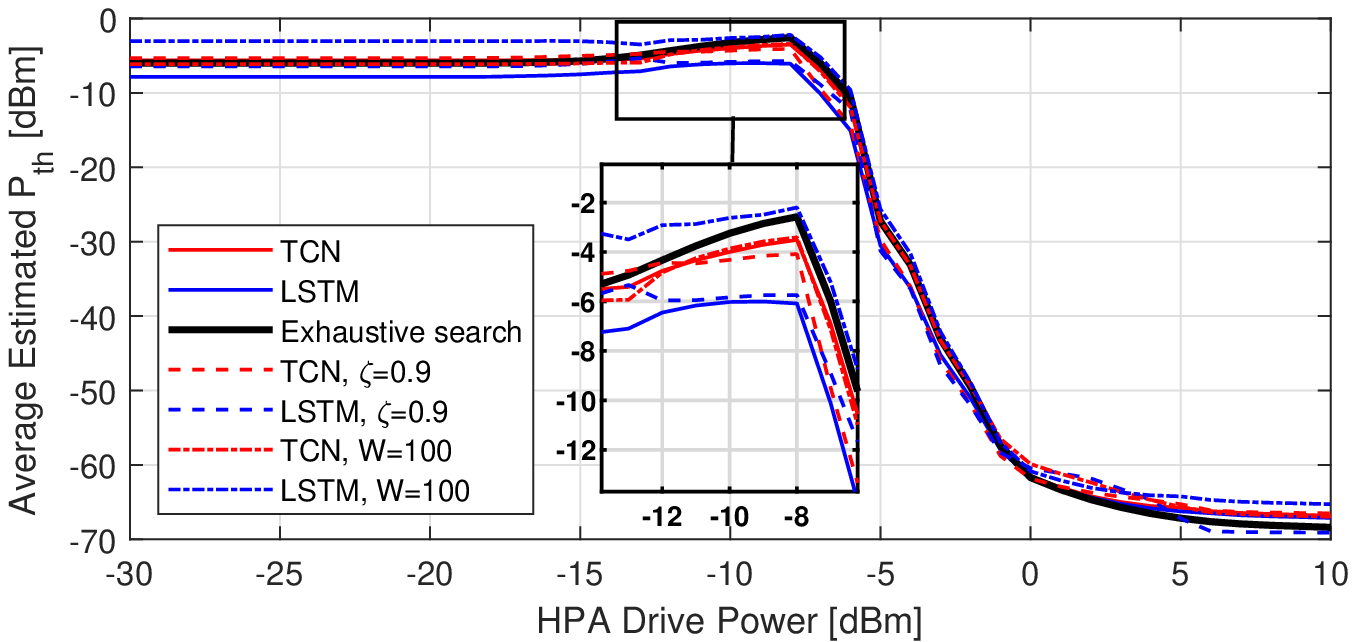}
}
\subfigure[]{
\includegraphics[width=3.6in]{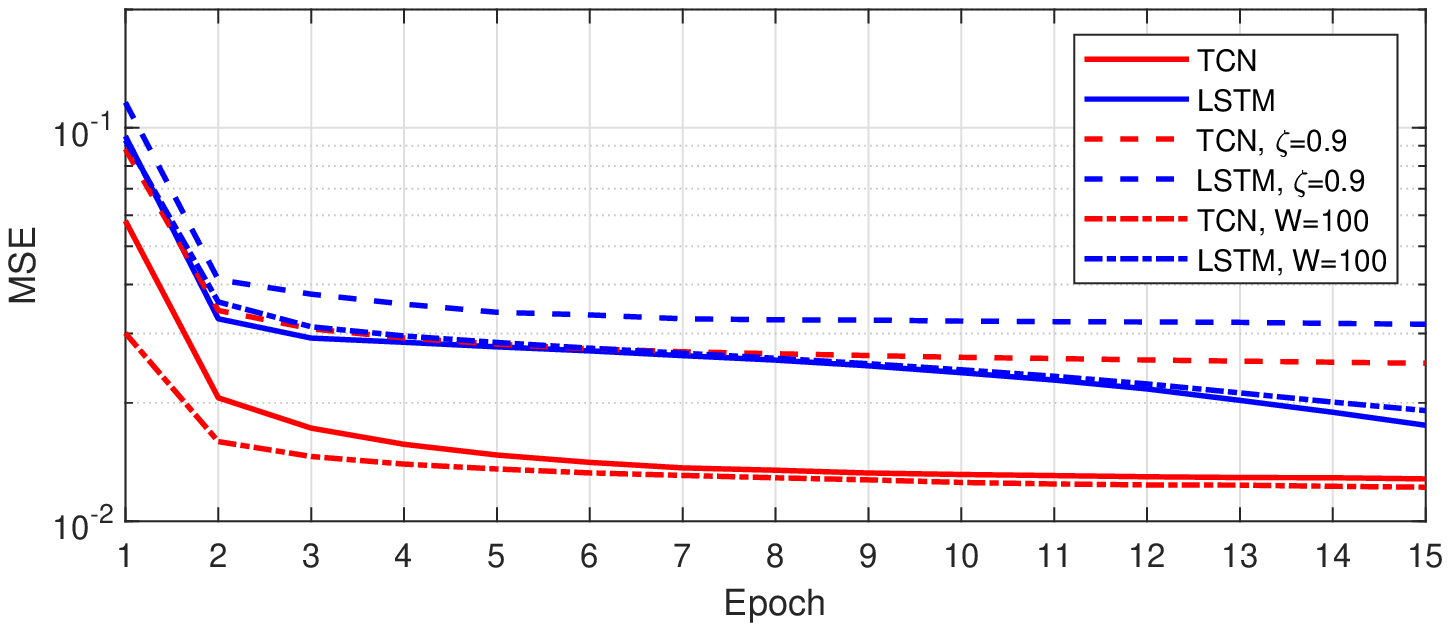}
}
\caption{Performance comparison of the TCN and LSTM RNN based adaptive control algorithms. (a) Average estimated $P\textsubscript{th}$ versus the HPA input drive power. (b) MSE versus the training epochs.}
\label{fig:Deep}
\end{figure}

In Fig.~\ref{fig:Deep}(a), the estimated $\hat{P}\textsubscript{th}$ using TCN and LSTM RNN is shown versus the HPA drive power. We notice that the optimal $P\textsubscript{th}$ varies with the HPA drive power to optimize the behavior of the unified SWIPT. At a low HPA drive power, the receiver cannot meet the energy-causality constraint due to the small received power. Also, the HPA nonlinearity does not largely affect the adaptive control of the unified SWIPT in this region. Therefore, the MS threshold is set to a constant value, namely the receiver RF-to-DC PCE crossover point of single tone/multi-tone to maximize the EH efficiency. As the HPA drive power increases, the energy-causality constraint forces the MS threshold $P\textsubscript{th}$ to be increased. This way the multi-tone mode is more likely selected to boost up the PCE and ID performance. At this region, the TCN shows better estimation performance compared to the LSTM, as shown in the subplot of Fig.~\ref{fig:Deep}(a). This result coincides well with the better sequence modeling performance of TCN discussed in \cite{SB, YC}. At a large HPA drive power, the multi-tone signal is largely distorted due to the HPA nonlinearity. Thus, the adaptive control algorithm drastically decreases the MS threshold, and the single tone mode is more likely selected. Overall, we confirm that the TCN and LSTM RNN of the proposed adaptive control algorithm can predict the average optimal MS threshold. 

Fig.~\ref{fig:Deep}(b) shows the MSE of the TCN and LSTM RNN based algorithms versus the training epochs. We see that the MSE decreases as the number of epochs increases, and the training of the deep networks proceeds well. For comparison, we also plotted some other cases; 1) the temporal correlation coefficient $\zeta$ is reduced, and 2) the input sequence length is large. The first case reflects more time-varying situations, e.g., higher mobility of devices. We observe that the converging MSE value increases because both the TCN and LSTM RNN have limitation in estimating $P\textsubscript{th}$ due to the reduced temporal channel correlation. But we still see that the TCN outperforms the LSTM RNN while both algorithms roughly following the general trend of $P\textsubscript{th}$ in Fig.~\ref{fig:Deep}(a).

Meanwhile, when the input sequence length is increased to $W=100$, the convergence of the TCN is improved compared to the LSTM RNN with no improvement. 
This is because the TCN can retain long memory more realistically compared to the LSTM RNN.
However, the results reveal that the TCN converges well regardless of the window size $W$, while the LSTM RNN still requires a large number of iteration epochs for convergence. Since a large sequence length increases the computational complexity per iteration in the training of the sequence modeling, it is desired to select a small window size (sequence length) with the TCN to boost the convergence speed of the proposed adaptive control algorithm. 

\begin{figure}
\centering
\includegraphics[width=3.6in]{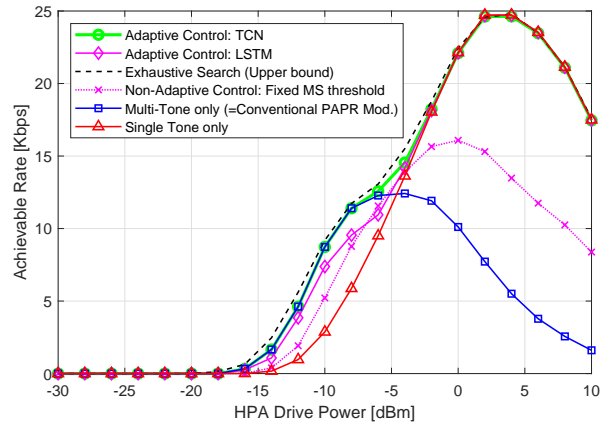}
\caption{Achievable rate of the unified SWIPT with TCN based adaptive control algorithm.}
\label{fig:Rate}
\end{figure}

In Fig.~\ref{fig:Rate}, we show the achievable rate versus the HPA drive power, subject to the self-powering condition. When the HPA drive power is very low, the ID of the unified SWIPT appears infeasible as the harvested power at the receiver cannot satisfy the self-powering condition. We see that the multi-tone mode achieves higher data rate in low-power region, whereas the single tone mode performs well in high-power region. However, the achievable rate is dramatically decreased at very high HPA drive power due to HPA nonlinearity. 

Compared to the SWIPT which utilizes single tone or multi-tone mode only, the unified SWIPT with exhaustive search or deep learning network achieves higher data rate through adaptive control. This is because the adaptive control algorithm updates $P\textsubscript{th}$ and switches the communication mode with optimum MS attributes considering both the received power feedback and HPA drive power, so as to maximize the achievable rate. Note that the exhaustive search requires huge computational complexity for real-time applications, thus it can be regarded as an ideal case of achieving the upper bound on the performance. We confirm that the proposed adaptive control algorithm with TCN produces the best performance compared to other baseline schemes. Furthermore, it offers a better achievable rate than the LSTM RNN because of the better estimation performance with TCN. 

Finally, we compare with the non-adaptive scheme with fixed MS threshold which is set to the receiver RF-to-DC PCE crossover point of single tone and multi-tone modes. We can observe that the performance is degraded compared to the adaptive control algorithms with TCN and LSTM RNN. Especially, the achievable rate is largely decreased at high HPA drive power as the fixed MS threshold is not optimized with regard to HPA nonlinearity. 
To the contrary, the adaptive control algorithms iteratively estimate $P\textsubscript{th}$ using the previous information. Therefore, the fine-tuning of MS threshold and MS attributes in the adaptive control algorithms results in enhanced performance. 

\section{Conclusions}
In this paper, we have proposed a novel unified signal and its architecture design for SWIPT with TCN based adaptive control algorithm. 
The unified signal and integrated receiver, which support both single tone and multi-tone signaling by adjusting only the power allocation ratio, were proposed and analyzed. In order to lower the computational burden of the receiver, which is the prerequisite for realizing {\it battery-free} IoT devices, the TCN based adaptive control algorithm was adopted, by which the transmitter can optimize the modulation index and the power allocation ratio in short-term scale while updating the MS threshold in long-term scale. To validate the system performance improvement, we have evaluated the achievable rate under the energy-causality constraint. It was confirmed that the proposed unified signaling and integrated receiver can produce higher achievable rate under the self-powering condition than the existing SWIPT. Therefore, the proposed unified SWIPT system with adaptive control at the transmitter side will be a promising enabler for self-powering low-power IoT devices {\it remotely}.

\appendices
\section{Proof of the CDF of PAPR modulation}\label{apd:CDF_PAPR}
First, we here omit the channel index $v$ for analytical simplicity. By conditioning on $|h|$, the CDF of $\text{PAPR}_i(N)$ is expressed as
\begin{equation}
\begin{aligned}
F_{i}(\gamma,N)&=\text{Pr}\, \Big\{\!\: \text{PAPR}_{i}(N)<\gamma \,\Big|\, |h| \Big\} \\
&=\text{Pr}\, \Big\{\!\: \!\max_{t}\, \big|y_{i}(t)\big|^2  < \gamma \mathcal P_i(\rho,N)  \,\Big|\, |h| \Big\}.
\end{aligned}
\end{equation}
Since $\{y\textsubscript{i}(t)\}$ are mutually independent over $t$, the CDF of the largest order statistic is derived as
\begin{equation}
F_i(\gamma,N)=\prod_{t} \!\:\text{Pr}\, \Big\{ \big|y_{i}(t)\big|^2  < \gamma \mathcal P_i(\rho,N) \,\Big|\, |h| \Big\}.
\end{equation}
Since $Y_i(t)$ is a deterministic signal for a given symbol, $y_i (t)\sim\mathcal{N}(Y_i(t),\sigma_{i}^2)$. Let us denote $X=|y_{i}(t)|^2/\sigma_{i}^2$, then $X$ is noncentral chi-squared random variable with one degree of freedom and noncentrality parameter $\lambda_i=[Y_i(t)/\sigma_{i}]^2$. Thus, the CDF is rewritten as
\begin{equation}
\label{eqn:CDF_exp}
\begin{aligned}
F_{i}(\gamma,N)&=\prod_{t} \!\:\text{Pr}\, \Bigg\{ X < \frac{\gamma}{\sigma_{i}^2} \mathcal P_i(\rho,N) \,\bigg|\, |h| \Bigg\} \\
&=\prod_{t} \!\:\Bigg\{ 1-\mathbb{E}_{|h|} \Big[Q_{1/2} \Big(\!\sqrt{\lambda_i},\sqrt{\nu_i}\,\Big) \Big] \Bigg\} 
\end{aligned}
\end{equation}
where $Q_{1/2} (\!\sqrt{\lambda_i},\sqrt{\nu_i})$ denotes the Marcum-Q function of order $1/2$ and $\nu_i=\gamma \mathcal P_i(\rho,N)/\sigma_{i}^2$. Unconditioning on $|h|$ using the channel PDF $f_{|h|}(z)$ completes the proof. 

For example, we take the expectation of (\ref{eqn:CDF_exp}) with respect to Rayleigh fading $h\sim\mathcal{CN}(0,\sigma_{h}^2)$, which yields
\begin{equation}
\label{eqn:CDF_PAPR}
\begin{aligned}
&F_{i}(\gamma,N)\\
&=\prod_{t} \Bigg\{1-\int_{0}^{\infty}\frac{z}{\sigma_h^2} \exp\left(-\frac{z^2}{2\sigma_h^2}\right) Q_{1/2} \Big(\!\sqrt{\lambda_i},\sqrt{\nu_i}\,\Big)\, dz \Bigg\}.
\end{aligned}
\end{equation}

\end{document}